\documentclass{emulateapj}
\shorttitle{Spectral Evolution of SN~2006gy}
\shortauthors{Smith et al.}
\begin{document}

\title{Spectral Evolution of the Extraordinary Type II\lowercase{n}
  Supernova 2006\lowercase{gy}}

\author{Nathan Smith\altaffilmark{1,2}, Ryan Chornock\altaffilmark{1},
  Jeffrey M.\ Silverman\altaffilmark{1}, Alexei V.\
  Filippenko\altaffilmark{1}, \& Ryan J.\ Foley\altaffilmark{1,3,4}}

\altaffiltext{1}{Department of Astronomy, University of California,
  Berkeley, CA 94720-3411.}
\altaffiltext{2}{Email: nathans@astro.berkeley.edu.}
\altaffiltext{3}{Center for Astrophysics, 60 Garden St., Cambridge,
  MA  02138.}
\altaffiltext{4}{Clay Fellow.}

\begin{abstract}

  We present a detailed analysis of the extremely luminous and
  long-lasting Type~IIn supernova (SN) 2006gy using spectra obtained
  between days 36 and 237 after explosion.  We derive the temporal
  evolution of the effective temperature, radius, blast-wave and
  SN-ejecta expansion speeds, and bolometric luminosity, as well as
  the progenitor wind density and total swept-up mass overtaken by the
  shock.  SN~2006gy can be interpreted in the context of shock
  interaction with a dense circumstellar medium (CSM), but with quite
  extreme values for the CSM mass of $\sim$20~M$_{\odot}$ and a SN
  explosion kinetic energy of at least 5$\times$10$^{51}$~erg.
  A key difference between SN~2006gy and other SNe~IIn is that, owing
  to its large amount of swept-up mass, the interaction region
  remained opaque much longer.  At early times, H$\alpha$
  emission-line widths suggest that the photosphere is ahead of the
  shock, and photons diffuse out through the opaque CSM shell.  The
  pivotal transition to optically thin emission begins to occur around
  day 110, when we start to see a decrease in the blackbody radius
  $R_{\rm BB}$ and strengthening tracers of the post-shock shell.  (A
  lingering puzzle, however, is that the late-time data require
  additional energy not traced by CSM-interaction diagnostics.)  From
  the evolution of pre-shock velocities, we deduce that the CSM was
  ejected by the progenitor star in a $\sim$10$^{49}$ erg precursor
  event $\sim$8 yr before the explosion.  The large CSM mass
  definitively rules out models involving stars with initial masses of
  $\lesssim$10 M$_{\odot}$.  If the pre-SN mass budget also includes
  the likely SN ejecta mass of 10--20 $M_{\odot}$ and the distant
  $>$10~M$_{\odot}$ shell inferred elsewhere for the infrared light
  echo, then even massive $M_{\rm ZAMS}$=30--40 M$_{\odot}$ progenitor
  stars are inadequate.  At roughly solar metallicity, substantial
  mass loss probably occurred during the star's life, so SN~2006gy's
  progenitor is more consistent with sequential giant luminous blue
  variable eruptions or pulsational pair-instability ejections in
  extremely massive stars with initial masses above 100 M$_{\odot}$.
  This requires significant revision to current paradigms of
  massive-star evolution.

\end{abstract}

\keywords{circumstellar matter --- stars: evolution --- supernovae:
individual (SN~2006gy)}

\section{INTRODUCTION}

The subclass of Type~IIn supernovae (SNe~IIn) provides a unique probe of
extreme pre-SN mass loss in massive stars.  The IIn designation
(Schlegel 1990; Filippenko 1997) refers to either (1) very narrow 
(widths of tens to a few hundred km s$^{-1}$) H emission lines that 
are thought to arise from photoionized pre-shock circumstellar gas
(e.g., Salamanca et al.\ 1998, 2002; Chugai et al.\ 2002), or (2)
intermediate-width (typically (1--5) $\times 10^3$ km s$^{-1}$) H lines
from post-shock gas accelerated as the SN ejecta collide with the
circumstellar medium (CSM), sometimes with extended wings broadened by 
electron scattering (Chugai \& Danziger 1994; Chugai 2001; Chugai et 
al.\ 2004; Dessart et al.\ 2009).

SNe~IIn show a huge range in bolometric luminosity and spectral
properties, presumably resulting from a wide range of pre-SN CSM
environments and intrinsic SN properties.  Their visual-wavelength
luminosity in excess of a normal Type II-plateau (II-P) SN is powered
largely by reprocessing SN ejecta kinetic energy through shocks
arising from CSM interaction (Chugai 1990; Chevalier 1977, 2003).
Depending on the strength of CSM interaction and optical depth of the
emitting region, SNe~IIn can potentially be luminous X-ray and radio
emitters for years after explosion, as in the cases of SNe~1988Z,
1986J, 1980K, 1978K, and others (e.g., Canizares et al.\ 1982;
Leibundgut et al.\ 1991; Ryder et al.\ 1992; Bregman \& Pildis 1992;
Van Dyk et al.\ 1993; Williams et al.\ 2002; Pooley et al.\ 2002;
Schlegel \& Petre 2006).

\begin{figure}   
\epsscale{0.99}
\plotone{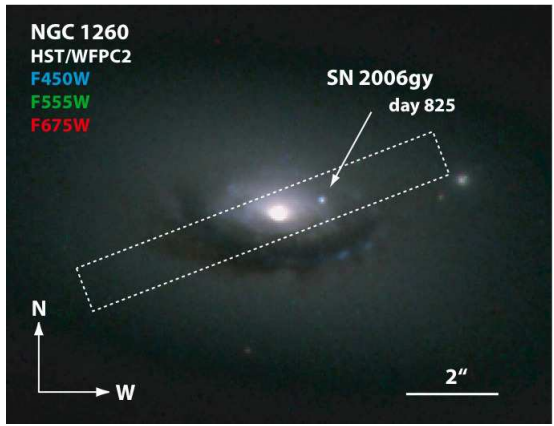}
\caption{Color {\it HST}/WFPC2 image of NGC~1260 and SN~2006gy
  obtained $\sim$2 yr after explosion (on day 825), with the F450W
  filter in blue, F555W in green, and F675W in red. (The acquisition,
  reduction, and analysis of these late-time {\it HST}/WFPC2 data are
  presented in a companion paper; Miller et al.\ 2009b.) At early
  epochs when the SN dominated the light, the spectroscopic slit
  aperture was oriented at the parallactic angle, while at later
  epochs the slit was oriented at a position angle of 285\arcdeg\ --
  291\arcdeg\ such that it passed through the SN and the galactic
  nucleus so that we could be sure of the supernova's position.  A
  typical slit aperture is shown.}
\label{fig:wfpc2}
\end{figure}

At the time of discovery, SN~2006gy was by far the most luminous known
SN~IIn.  It exploded in the peculiar S0/Sa galaxy NGC~1260
(Figure~\ref{fig:wfpc2}), and the early data and significance of this
SN were first discussed by Ofek et al.\ (2007) and Smith et al.\
(2007; Paper~I hereafter).  SN~2006gy was extraordinary in that it had
a peak luminosity more than 100 times higher than that of a normal
SN~II-P, it had a remarkably slow rise time of $\sim$70 days, the
light-curve shape was unusually rounded for a SN~IIn (Paper~I), and
the CSM environment was extremely dense and extended.  Analysis of the
light-curve rise time, including pre-discovery photometry, implied a
rough explosion date of 2006 Aug.\ 20 (Paper~I; UT dates are used
throughout this paper), which we shall adopt here.  SN~2006gy was
located only 1\arcsec\ ($\sim$300 pc) from the active nucleus of
NGC~1260, which has provided enduring difficulty in studying the
object as it fades, and the SN appeared to be located along a dust
lane with superposed emission from H~{\sc ii} regions (Ofek et al.\
2007; Paper~I; Smith et al.\ 2008b; Figure~\ref{fig:wfpc2}).
NGC~1260's star-formation rate of $\sim$1.2 M$_{\odot}$ yr$^{-1}$
indicated by its infrared (IR) emission makes it plausible that the
progenitor of SN~2006gy was a massive star associated with a young
stellar population (Ofek et al.\ 2007; Paper~I). The host galaxy's
average metallicity is 0.6--0.7 $Z_{\odot}$, and SN~2006gy occurred
near the nucleus, so its environment was {\it not} low metallicity.
This influences its pre-SN mass loss and any models for the progenitor
star's evolution.


The high luminosity and large total radiated energy of SN~2006gy --
exceeding 10$^{51}$ erg in visual light alone -- raised new questions
about the engine that powers the luminosity, which we have discussed
in detail in Paper~I.  In brief, the two leading candidates are CSM
interaction, motivated by the Type IIn spectrum, or radioactive decay
from several M$_{\odot}$ of $^{56}$Ni generated in a pair-instability
supernova (PISN).  Neither hypothesis matches straightforward
expectations.  The CSM interaction hypothesis is complicated by the
astonishingly large required mass (10--20 M$_{\odot}$) in the CSM
environment, which in turn seems to require an explosive or impulsive
shell ejection to occur within a few years immediately preceding the
SN (see Paper~I; Smith \& McCray 2007; Woosley et al.\ 2007).  Key
shock diagnostics such as H$\alpha$ and X-ray emission in SN~2006gy,
however, are far weaker than naive expectations for CSM interaction.

On the other hand, the PISN hypothesis is complicated by the fact that
true PISNe (Barkat et al.\ 1967; Rakavy \& Shaviv 1967; Bond et al.\
1984; Heger \& Woosley 2002) are only expected for extremely massive
progenitor stars at low metallicity where mass loss is thought to be
unimportant (e.g., Heger et al.\ 2003), and the lethargic light-curve 
evolution of SN 2006gy was not slow enough to match predictions for PISNe
(Scannapieco et al.\ 2005).  Effects of close binary evolution, such
as common-envelope ejection or mergers, have also been discussed (Ofek
et al.\ 2007; Portegies Zwart \& van den Heuvel 2007), although these
do not address the physical origin of the high luminosity itself.
Agnoletto et al.\ (2009) invoked a combination of these extraordinary
conditions, placing somewhat less extreme demands than for any single
mechanism alone: radioactive decay from 1--3 M$_{\odot}$ of $^{56}$Ni
plus extraordinarily strong CSM interaction with a 6--10 M$_{\odot}$
pre-SN envelope.  In any case, SN~2006gy stretched the limits for
plausible physical explanations of its energy source.

\begin{deluxetable*}{lclcccccc}\tabletypesize{\scriptsize}
\tablecaption{Spectroscopic Observations of SN~2006\lowercase{gy}}
\tablewidth{0pt}
\tablehead{
  \colhead{UT Date} &\colhead{Day\tablenotemark{a}} &\colhead{Instrument}
  &\colhead{Range\tablenotemark{b}} &\colhead{$\lambda$/$\Delta\lambda$} 
  &\colhead{T$_{\rm BB}$(cont.)\tablenotemark{c}} &\colhead{EW(H$\alpha$)}  &\colhead{$F$(H$\alpha$)} 
  &\colhead{H$\alpha$/H$\beta$\tablenotemark{d}} \\
  \colhead{} &\colhead{} &\colhead{} &\colhead{(\AA)} &\colhead{} &\colhead{(10$^3$ K)} 
  &\colhead{(\AA)}  &\colhead{(erg s$^{-1}$ cm$^{-2}$)} &\colhead{} }
\startdata

2006~Sep.~25  &36  &Kast/Lick    &3500--9600  &700  &12.0(3.0) &$-$56.1(3) &13.5(0.8)$\times$10$^{-14}$ &2.9(0.2) \\
2006~Oct.~24  &65  &Kast/Lick    &3500--9600  &700  &12.0(0.8) &$-$28.0(2) &10.4(0.6)$\times$10$^{-14}$ &2.9(0.3) \\
2006~Oct.~30  &71  &Kast/Lick    &3500--9600  &700  &11.0(0.7) &$-$23.5(2) &8.7(0.5)$\times$10$^{-14}$  &3.2(0.3) \\
2006~Nov.~20  &92  &Kast/Lick    &5450--6800  &1000 &8.5(0.9)  &$-$18.0(2) &5.4(0.4)$\times$10$^{-14}$  &\nodata  \\
2006~Nov.~21  &93  &LRIS/Keck    &3500--9100  &1000  &9.00(0.3) &$-$22.0(1) &6.9(0.3)$\times$10$^{-14}$  &6.8(2)   \\
2006~Nov.~24  &96  &DEIMOS/Keck  &4500--7100  &8000 &8.80(0.4) &$-$27.7(1) &7.8(0.2)$\times$10$^{-14}$  &6.5(2)   \\
2006~Dec.~20  &122 &Kast/Lick    &3500--9600  &700  &7.10(0.3) &$-$57.4(4) &10.0(0.4)$\times$10$^{-14}$ &$\ga$50  \\
2006~Dec.~23  &125 &DEIMOS/Keck  &4500--7100  &8000 &7.05(0.4) &$-$61.4(1) &10.1(0.2)$\times$10$^{-14}$ &45(15)   \\
2006~Dec.~25  &127 &LRIS-P/Keck  &3500--9050  &1000  &7.05(0.3) &$-$64.8(2) &10.5(0.4)$\times$10$^{-14}$ &65(20)   \\
2007~Jan.~13  &145 &LRIS/Keck    &3500--9050  &1000  &6.9(0.8)  &$-$104(4)  &11.9(0.4)$\times$10$^{-14}$ &128(40)  \\
2007~Jan.~21  &153 &LRIS-P/Keck  &3500--9050  &1000  &6.60(0.3) &$-$133(5)  &9.17(0.3)$\times$10$^{-14}$ &105(30)  \\
2007~Jan.~22  &154 &DEIMOS/Keck  &4500--7100  &8000 &6.2(0.9)  &$-$114(2)  &9.14(0.2)$\times$10$^{-14}$ &110(30)  \\
2007~Jan.~26  &158 &Kast/Lick    &4200--9600  &700  &6.50(0.4) &$-$108(6)  &8.1(0.4)$\times$10$^{-14}$  &(abs.)   \\
2007~Feb.~14  &177 &LRIS-P/Keck  &3500--9050  &1000  &6.55(0.3) &$-$103(4)  &5.15(0.2)$\times$10$^{-14}$ &100(30)  \\
2007~Feb.~16  &179 &DEIMOS/Keck  &4500--7100  &8000 &6.45(0.7) &$-$115(3)  &5.62(0.2)$\times$10$^{-14}$ &92(20)   \\
2007~Mar.~18  &209 &LRIS-P/Keck  &3500--9050  &1000  &6.45(0.3) &$-$67.9(4) &2.4(0.2)$\times$10$^{-14}$  &78(15)   \\
2007~Apr.~15  &237 &LRIS-P/Keck  &3500--9050  &1000  &5.90(0.3) &$-$33.5(4) &1.1(0.2)$\times$10$^{-14}$  &\nodata  \\
2007~Oct.~16  &421 &LRIS-P/Keck  &3500--9050  &1000  &\nodata   &\nodata    &\nodata                    &\nodata  \\

\enddata
\tablenotetext{a}{Days since 2006 Aug. 20, the presumed time of
  explosion adopted from Paper~I.  This is $\sim$29 days before
  discovery.}  
\tablenotetext{b}{Spectral range in rest wavelength.}
\tablenotetext{c}{Approximate blackbody temperature indicated by the
  continuum shape in dereddened visual-wavelength spectra (see
  Fig.~\ref{fig:dered}). Here and elsewhere, uncertainties are given
  in parentheses. The temperature derived from the continuum
  shape for day 36 is less certain for reasons described in the text.}
\tablenotetext{d}{The H$\alpha$/H$\beta$ emission flux ratio measured 
  after the spectra are dereddened.  Many of these values are highly 
  uncertain because at later times H$\beta$ emission is weak, is 
  blended with Fe~{\sc ii} lines, and is affected by deep absorption.}
\end{deluxetable*}


While the PISN hypothesis has been difficult to disprove, there is a
potential way to reconcile the CSM interaction model with the apparent
difficulties posed by observations. Smith \& McCray (2007) proposed
that with the high optical depths that might occur when a blast wave
overtakes an opaque pre-SN shell, radiative diffusion may be
important, so that X-ray and H$\alpha$ emission would be weak compared
to other SNe~IIn.  In this model, the weak H$\alpha$ and unabsorbed
soft X-rays occur after the blast wave exits the opaque shell and
encounters additional CSM downstream, while radiation continues to
diffuse out from shock-deposited thermal energy in the slowly
expanding opaque shell.  Woosley et al.\ (2007) considered a similar
idea, but proposed that the pre-SN ejection in question was triggered
by the pulsational pair instability.  This occurs as a result of the
same pair instability that leads to a PISN, but the explosion is
insufficient to fully unbind the star, resulting in a lower-energy
ejection of only the outer envelope instead (Heger \& Woosley 2002;
Heger et al.\ 2003).  In their model, this occurred in the late
nuclear burning phases of a star with an initial mass of 110
M$_{\odot}$ and with lower mass-loss rates than have traditionally
been adopted in stellar evolution calculations for stars at solar
metallicity.

The idea that a very massive star retains a massive H envelope until
core collapse, while experiencing eruptive mass ejection analogous to
giant eruptions of luminous blue variables (LBVs) like $\eta$ Carinae,
has interesting implications for understanding massive-star evolution,
as discussed elsewhere (see Paper~I; Smith \& Owocki 2006; Gal-Yam et
al.\ 2007).  Although it defies current paradigms of stellar evolution
theory, the LBV/SN~IIn connection has recently been reinforced by the
direct detection of a progenitor of the Type IIn SN~2005gl, which
resembled a very luminous LBV (Gal-Yam \& Leonard 2009).
Nevertheless, the true nature of the progenitor of SN~2006gy remains
uncertain.


It was hoped that late-time observations could settle these mysteries
about the power source for the tremendous luminosity of SN 2006gy.
Smith et al.\ (2008b) reported optical and near-IR photometry and
spectroscopy obtained after one year.  The SN was below detection
limits at visual wavelengths, but in the near-IR it was still at least
as luminous as the peak of a normal SN~II-P.  The late-time spectrum
revealed no clear detection of features associated with SN~2006gy,
although the object is only 1\arcsec\ from the bright nucleus as noted
above.  However, these spectra yielded upper limits to any late-time
post-shock H$\alpha$ emission of $\la$10$^{39}$ erg s$^{-1}$, which is
a factor of $\sim$400 lower than for a less luminous SN~IIn such as
SN~1988Z (e.g., Aretxaga et al.\ 1999) at a comparable epoch.  (The
luminous SN~IIn 2006tf, by contrast, showed strong and easily detected
broad H$\alpha$ at late times; Smith et al.\ 2008a.)  If CSM
interaction were an important engine at late times, one would also
expect strong X-ray emission or radio emission, which was not the case
(Bietenholz \& Bartel 2007; Smith et al.\ 2008b).  Therefore, Smith et
al.\ (2008b) ruled out continuing CSM interaction as the late-time
luminosity source, and proposed either radioactive decay obscured by
dust or, more likely, an IR echo from a distant dust shell.


Subsequently, Agnoletto et al.\ (2009) and Kawabata et al.\ (2009)
also presented late-time optical and IR photometry of SN 2006gy,
including detections and limits consistent with Smith et al.\ (2008b).
Kawabata et al.\ claimed a detection of late-time ($>$1 yr) spectral
features associated with SN~2006gy, although some of the lines are
questionable given their peculiar or unknown identifications. Their
estimate of the late-time H$\alpha$ flux of $\sim$10$^{39}$ erg
s$^{-1}$ agrees with the upper limits of Smith et al.\ (2008b),
confirming that the late-time luminosity cannot be powered by CSM
interaction.  Kawabata et al.\ proposed that their late-time
photometric detection was consistent with a radioactive decay tail
from 15~M$_{\odot}$ of $^{56}$Ni (Nomoto et al.\ 2007) if
$\gamma$-rays from the decay are not fully trapped.  A light echo from
a massive CSM shell ejected $\sim$1500 yr before core collapse, as
proposed by Smith et al.\ (2008b), is the favored interpretation for
the late-time IR luminosity in view of more recent evidence, however.
In a companion paper, we discuss the IR-echo hypothesis in more detail
(Miller et al.\ 2009b).


Although SN~2006gy had the hghest peak luminosity of any known SN when
it was discovered (Ofek et al.\ 2007; Smith et al.\ 2007), subsequent
discoveries of SN~2005ap\footnote{Although the explosion of SN~2005ap
  obviously occurred before SN~2006gy, its distance and high
  luminosity were recognized later (Quimby et al.\ 2007).} (Quimby et
al.\ 2007) and SN~2008es (Gezari et al.\ 2009; Miller et al.\ 2009a)
have raised the bar, with peak luminosities almost a factor of 2
higher.  For all three events, the integrated radiated energy in the
first few months was of order 10$^{51}$ erg, and
$\sim$2$\times$10$^{51}$ erg in the case of SN~2006gy.  These objects
therefore present a new and fundamental challenge to SN physics.  With
a peak luminosity $\sim$1 mag fainter than that of SN~2006gy, the Type
IIn SN~2006tf seems to mark the upper bound of what can be achieved
with standard CSM interaction because of optical-depth limitations
(see Smith et al.\ 2008a).  Standard CSM interaction, taken here to
mean that the observed luminosity arises from direct cooling by the
radiative post-shock gas, therefore has difficulty accounting for the
higher luminosities of SNe~2006gy, 2005ap, and 2008es.  This leads one
to suspect that high optical depths and consequent delayed effects of
radiative diffusion play an essential role (Smith \& McCray 2007).

We have the opportunity to focus in detail on SN~2006gy because it was
closer ($d = 73.1$ Mpc) than either SN~2005ap ($d \approx 1200$ Mpc)
or SN~2008es ($\sim$900 Mpc), so available spectra of SN~2006gy are of
higher quality.  A consistent theme throughout this paper will be that
although there are clear signs of CSM interaction in the spectrum, the
level of CSM interaction one infers from spectral diagnostics seems
vastly insufficient to power the visual-wavelength continuum
luminosity of SN~2006gy.  This amplifies comments we made in Paper~I
regarding its relatively weak H$\alpha$ and X-ray emission.  The
insufficient level of CSM interaction in SN~2006gy is particularly
relevant to SN~2005ap and SN~2008es, since both of those events
appeared to be Type II-linear (II-L) events, lacking any narrow
H$\alpha$ emission indicative of the CSM interaction that one sees in
SNe~IIn.  All other top contenders with peak $M_R$ brighter than $-$20
mag were of Type~IIn.

In Paper~I, we considered basic implications of the energy budget of
SN~2006gy, while Smith \& McCray (2007) and Smith et al.\ (2008b) have
discussed additional aspects of the evolving light curve and late-time
photometry.  Independent observations of SN~2006gy have been discussed
by Ofek et al.\ (2007), Kawabata et al.\ (2009), and Agnoletto et al.\
(2009).  In this paper we undertake a more detailed analysis of our
multi-epoch spectra, concentrating on the main light-curve peak when
SN~2006gy released most of its radiated energy.  We adopt many of the
same basic properties as in Paper~I: $d = 73.1$ Mpc, $z = 0.0179$, and
$A_R = 1.68$ mag.  We present the spectral observations in \S 2
(including additional spectra not published in Paper~I), and we
examine the main results of these spectra in \S 3.  In \S 4 we
synthesize the observations into a general scenario to explain the SN,
and we also discuss implications for SNe and massive-star evolution.

\begin{figure*}[!ht]   
\epsscale{0.9}
\plotone{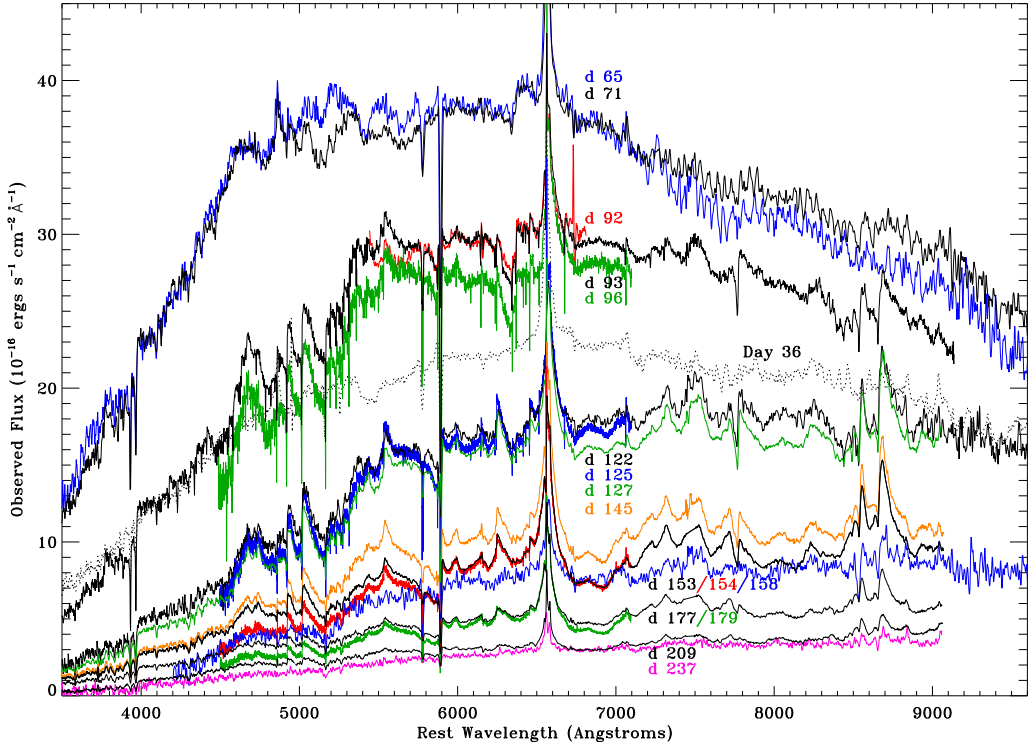}
\caption{Observed spectra of SN~2006gy during the main light-curve
  peak on the days indicated (see Table 1), plotted on a linear
  intensity scale.  These have not been corrected for any extinction.
  The dashed line is the day 36 spectrum, while the SN was still
  rising.}
\label{fig:flam}
\end{figure*}

\begin{figure}   
\epsscale{0.95}
\plotone{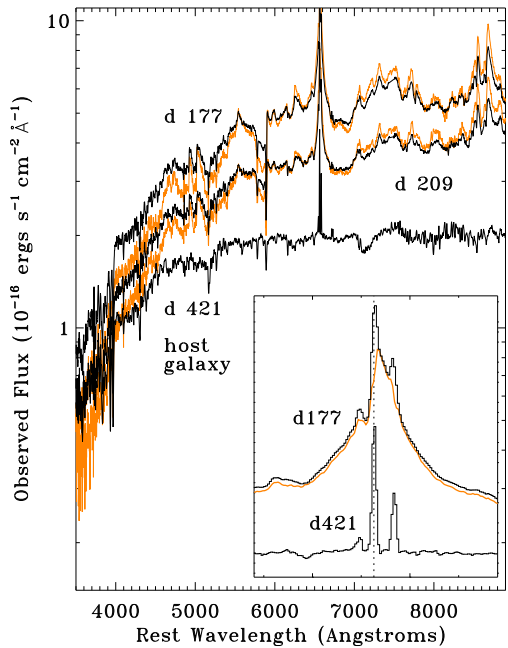}
\caption{Late-time spectra of SN~2006gy illustrating background-subtraction 
  issues.  The day 421 spectrum is the extracted flux at
  the position of SN~2006gy at a time when the SN has faded beyond
  detectability.  The day 177 and 209 spectra in black are the same as
  in other figures, where we have performed background subtraction by
  sampling the emission along the slit on each side of the SN.  The
  gray ({\it orange in the online edition}) spectra are for the same
  dates, but where we have subtracted an additional amount of
  background-galaxy light using the day 421 spectrum (scaled to the
  same red flux level).  These represent the most host-galaxy light
  that can be subtracted without causing erroneous narrow absorption
  at the position of nebular [N~{\sc ii}] lines (see \S 2.2).  The
  inset shows a detail of the spectral region around H$\alpha$ and
  [N~{\sc ii}].}
\label{fig:host}
\end{figure}

\begin{figure*}
\epsscale{0.9}
\plotone{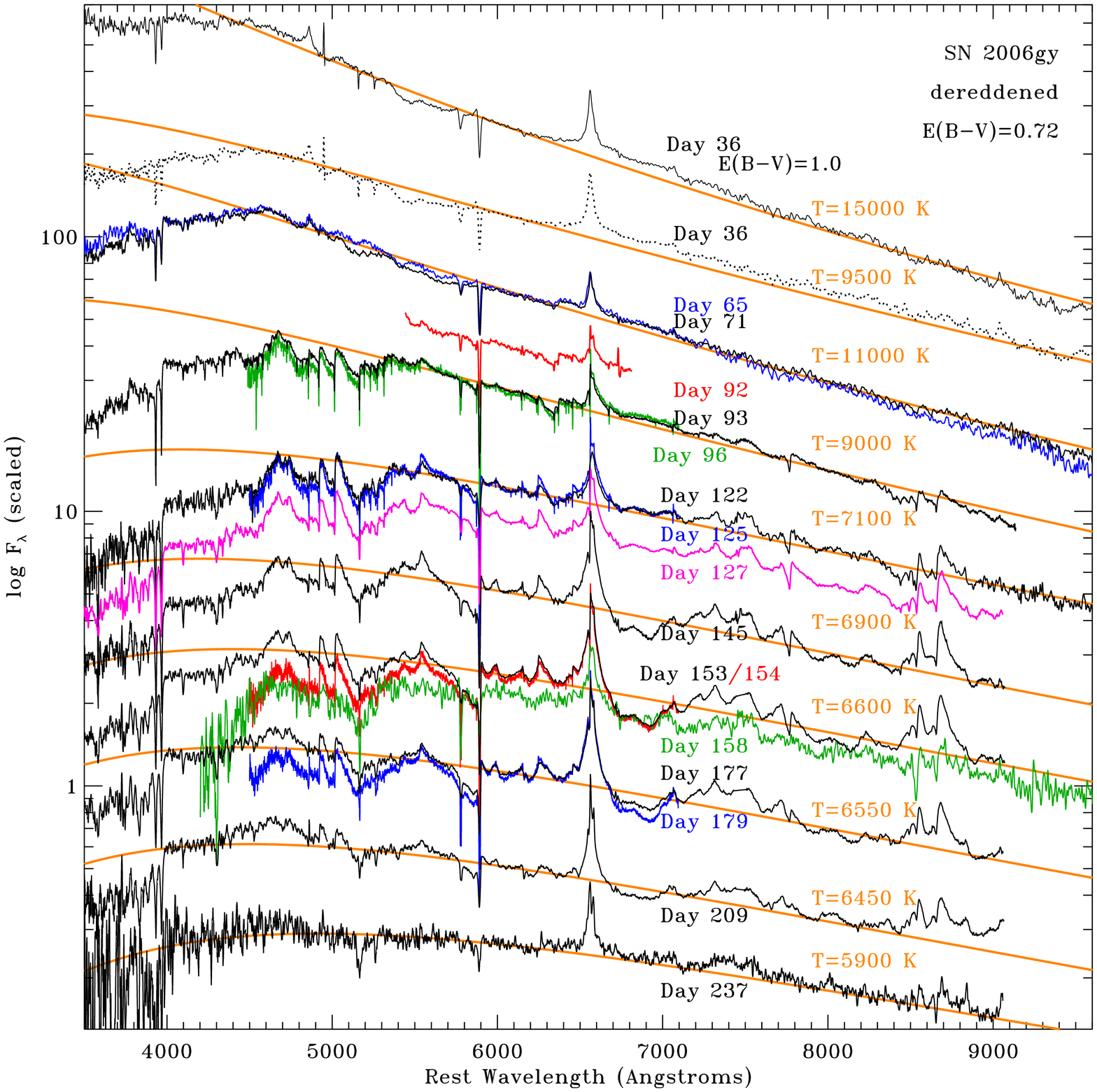}
\caption{Same as Figure~\ref{fig:flam}, but dereddened, plotted on a
  log scale, and scaled arbitrarily for display.  The spectra have
  been dereddened by correcting for Galactic extinction of $A_R = 0.43$
  mag and for the additional local extinction of $A_R = 1.25$ (Paper I),
  for a total $E(B-V) = 0.72$ mag.  The day 36 spectrum is plotted twice:
  once with the same reddening correction as at other epochs (dotted) and
  once with a larger reddening correction (solid black).
  Representative blackbody temperatures are shown for comparison in
  orange.}
\label{fig:dered}
\end{figure*}

\section{OBSERVATIONS}

\subsection{Lick and Keck Spectroscopy}

We obtained visual-wavelength spectra of SN~2006gy at several times
during the first year after explosion, as well as at one epoch more
than 1~yr after discovery.  Table~1 summarizes the observational
parameters for each epoch.  These were obtained at the Keck
Observatory using the low-resolution imaging spectrometer (LRIS; Oke
et al.\ 1995) and the Deep Imaging Multi-Object Spectrograph (DEIMOS;
Faber et al.\ 2003).  We also obtained spectra using the Kast double
spectrograph (Miller \& Stone 1993) mounted on the Lick Observatory
3-m Shane telescope.  The spectra at early times were all obtained
with the long slit oriented at the parallactic angle (Filippenko 1982)
to minimize losses caused by atmospheric dispersion, while later
spectra when the SN faded were oriented with the slit passing through
the nucleus of NGC~1260 (see Figure~\ref{fig:wfpc2}).

The spectra were reduced using standard techniques (e.g., Foley et al.
2003). Our spectra on days 36, 71, and 96 were already presented in
Paper~I, and additional details of the data-reduction steps can be
found there.  The spectra were corrected for atmospheric extinction
(Bessell 1999; Matheson et al.\ 2000) using standard stars observed at
an airmass similar to that of the SN and then flux calibrated using
our photometry of SN~2006gy.  When correcting the spectra for
extinction, we adopt a host-galaxy extinction of $A_R = 1.25 \pm 0.25$
mag ($A_V = 1.67$ mag; $E(B-V)$=0.54 mag) for SN~2006gy in addition to
the Galactic extinction of $A_R = 0.43$ mag ($A_V = 0.57$ mag;
$E(B-V)$=0.18 mag), and we use $R = 3.1$ and the reddening curve of
Cardelli et al.\ (1989), as described in Paper~I.  As noted there, the
Na~{\sc i}~D absorption in the spectrum of SN~2006gy may suggest even
higher reddening than we have assumed here, so our luminosity
estimates for SN~2006gy are conservative.  The observed and
flux-calibrated spectra of SN~2006gy are shown in
Figure~\ref{fig:flam}. In all of the spectra presented here, the
observed wavelengths have been corrected to the rest-frame wavelengths
using redshift $z = 0.0179$ (see Paper~I).


\subsection{Difficulties Posed by the Environment}

Photometric and spectroscopic observations of SN~2006gy are
challenging because the SN is located only $\sim$1\arcsec\ from the
bright galactic nucleus of NGC~1260 (Figure~\ref{fig:wfpc2}). This is
especially true at late times when the SN has faded; during the main
light-curve peak, SN~2006gy was much stronger than shown in
Figure~\ref{fig:wfpc2}, dominating the surrounding flux (see Fig.~1 in
Paper~I).  There is a steep intensity gradient in the underlying
galaxy light that includes a dust lane and velocity-dependent emission
lines from H~{\sc ii} regions that follow the rotation curve of the
galaxy (Figure~\ref{fig:wfpc2}; Paper~I; Ofek et al.\ 2007; Smith et
al.\ 2008b).  To subtract host-galaxy light, we sampled emission along
the slit immediately on either side of SN~2006gy (an example slit is
shown in Figure~\ref{fig:wfpc2}).  We checked carefully to make sure
that the sky-subtraction procedure did not artificially introduce
false narrow absorption components in this procedure.  We determined
that such oversubtraction did not affect the data; H, He~{\sc i}, and
other lines show similar narrow blueshifted absorption profiles, even
though only H$\alpha$ and [N~{\sc ii}] are strong in the background,
and the observed narrow absorption components weaken systematically
with time as the SN fades.


To check the uncertainty introduced by the difficult host-galaxy
subtraction on such a strong intensity gradient near the nucleus, we
compare (Figure~\ref{fig:host}) two relatively late-time spectra of
SN~2006gy on days 177 and 209 to an extracted spectrum of the
background galaxy at the same position.  This background spectrum is
from day 421 when the SN is no longer detected in our data.  The flat
continuum level here has been scaled to the background intensity of
$\sim$2$ \times 10^{-16}$ erg~s$^{-1}$~cm$^{-2}$~\AA$^{-1}$~arcsec$^2$
measured in the WFPC2 image.  At the shortest wavelengths, the
background flux that would contaminate a $\sim$1\arcsec\ point source
is comparable to the SN flux at late times, so careful subtraction is
important. Indeed, below $\lambda \approx 4500$~\AA, late-time spectra
of the SN are nearly identical to the background, so we have some
concern that even our background-subtracted spectra may still include
substantial galaxy light in the blue.

In order to evaluate the level of remaining host-galaxy contamination
even after this baseline subtraction (which may be imperfect because
of differing slit-width/seeing combinations), we experimented with
subtracting additional galaxy light using the day 421 spectrum to
represent the background at the SN position.  The day 421 spectrum
shows strong, narrow H$\alpha$ and [N~{\sc ii}] emission lines from
H~{\sc ii} regions.  The most additional galaxy light one can subtract
is the amount which causes the nebular [N~{\sc ii}] line to disappear
in the SN spectrum; subtracting more than this would cause a spurious
absorption feature at 6583~\AA.  Using this criterion, the days 177
and 209 spectra with the most possible host-galaxy light subtracted
are shown in Figure~\ref{fig:host} with gray tracings ({\it orange in
  the online edition}).  The resulting spectra have weaker flux and
steeper continua at the shortest wavelengths.  Longward of 5000 \AA,
the contrast of some spectral features is stronger but the spectra are
otherwise very similar.  We suspect, however, that this may be an
oversubtraction, because the method assumes that the SN has no
intrinsic nebular emission from H$\alpha$ and [N~{\sc ii}], whereas
earlier spectra show variable intrinsic fluxes of these narrow lines,
and the day 179 DEIMOS spectrum with high resolution shows a resolved
narrow H$\alpha$ line profile consistent with earlier spectra when the
line was stronger.

Therefore, throughout this paper we adopt the original method, used
consistently for all observed epochs, of subtracting the background by
sampling emission along the slit on either side of the SN.  The
potentially oversubtracted spectra in Figure~\ref{fig:host} are
useful, however, to remind us of the level of uncertainty introduced
by possible remaining galaxy light in the data.  The differences
introduced by the additional subtraction are not severe -- they are
comparable, for example, to the differences between LRIS and DEIMOS
spectra at a given epoch (see Figure~\ref{fig:dered}).  If this
contamination were as severe as the worst case represented in
Figure~\ref{fig:host}, then the way that the additional subtraction of
host-galaxy light would alter our analysis is that at late times after
day 150 (and {\it only} at late times), we would derive a continuum
temperature (\S 3.2) that is $\sim$5--10\% lower and an H$\alpha$
equivalent width (\S 3.4.1) that is a few percent larger.  Such
differences are consistent with our estimated uncertainties, so it
would not significantly affect our analysis.  The subtraction issues
discussed here would not alter the measurements of line flux and
luminosity, and our conclusions do not rely on the flux at the
shortest wavelengths.

The inset of Figure~\ref{fig:host} shows in detail the spectral region
near H$\alpha$.  In our late-time spectrum on day 421, we see no
evidence for broad post-shock H$\alpha$ emission that might arise from
late-time CSM interaction, as we reported earlier (Smith et al.\
2008b).  The continuum underlying the narrow H$\alpha$ and [N~{\sc
  ii}] emission lines is flat, in stark contrast to earlier epochs
when broad H$\alpha$ was present.  Kawabata et al.\ (2009) reported
broad H$\alpha$ emission
on day 394, estimated from a raised flux level in the gap between
H$\alpha$ and [N~{\sc ii}] $\lambda$6583.  Figure~\ref{fig:host} shows
that at higher spectral resolution, this feature is not seen in our
data on day 421.  It therefore seems possible that the excess flux
reported by Kawabata et al.\ (2009) may have been a blend of the wings
of narrow H$\alpha$ and [N~{\sc ii}] emission lines in their
lower-resolution spectra, and may not have been attributable to the
SN.


In general, our flux-calibrated spectra give reliable and 
self-consistent results.  Compare, for example, the pair of spectra in
Figures~\ref{fig:flam} and \ref{fig:dered} on days 93 and 96, or the
pair on days 122 and 125.  These were obtained only a few days apart
with different instruments on different telescopes, but the flux
calibration, continuum slopes, and relative line strengths agree
remarkably well except perhaps for the strengths of some of the
narrowest emission or absorption features that are affected by the
different instrument spectral resolutions.

We did encounter a potential problem in calibrating our spectra,
however, and that was for the 2006 Sep.\ 25 (day 36) spectrum
obtained at Lick Observatory.  This was our first spectrum of
SN~2006gy, calibrated with standard stars that differed from those of 
all other epochs; more importantly, the blue-channel flatfield lamp on 
the Kast spectrograph was not working on this night, so we had to use 
a flatfield observation from a different observing run.  These issues
compromise the reliability of the flux calibration and continuum slope
on day 36, which is important for the resulting continuum temperature
that is derived on this epoch.  The red channel should not have been
affected by the problematic blue-channel flatfield lamp, and this set
of problems did not affect the remaining epochs of spectra.  We
therefore regard the blue continuum shape on day 36 with caution when
deriving physical quantities in our analysis. On the other hand, the 
strange continuum shape may be real and caused by unusual physical
circumstances; it could result from higher extinction due to
time-dependent circumstellar reddening, for example, since at that
early time the optical depth was very high, as we discuss below.  The
dereddened spectra for all epochs are shown in Figure~\ref{fig:dered}.
We adopted a correction of $E(B-V) = 0.723$ mag (see Paper I), although
for day 36 we also show an additional reddening correction of
$E(B-V) = 1.0$ mag.

\begin{figure*}
\epsscale{0.95}
\plotone{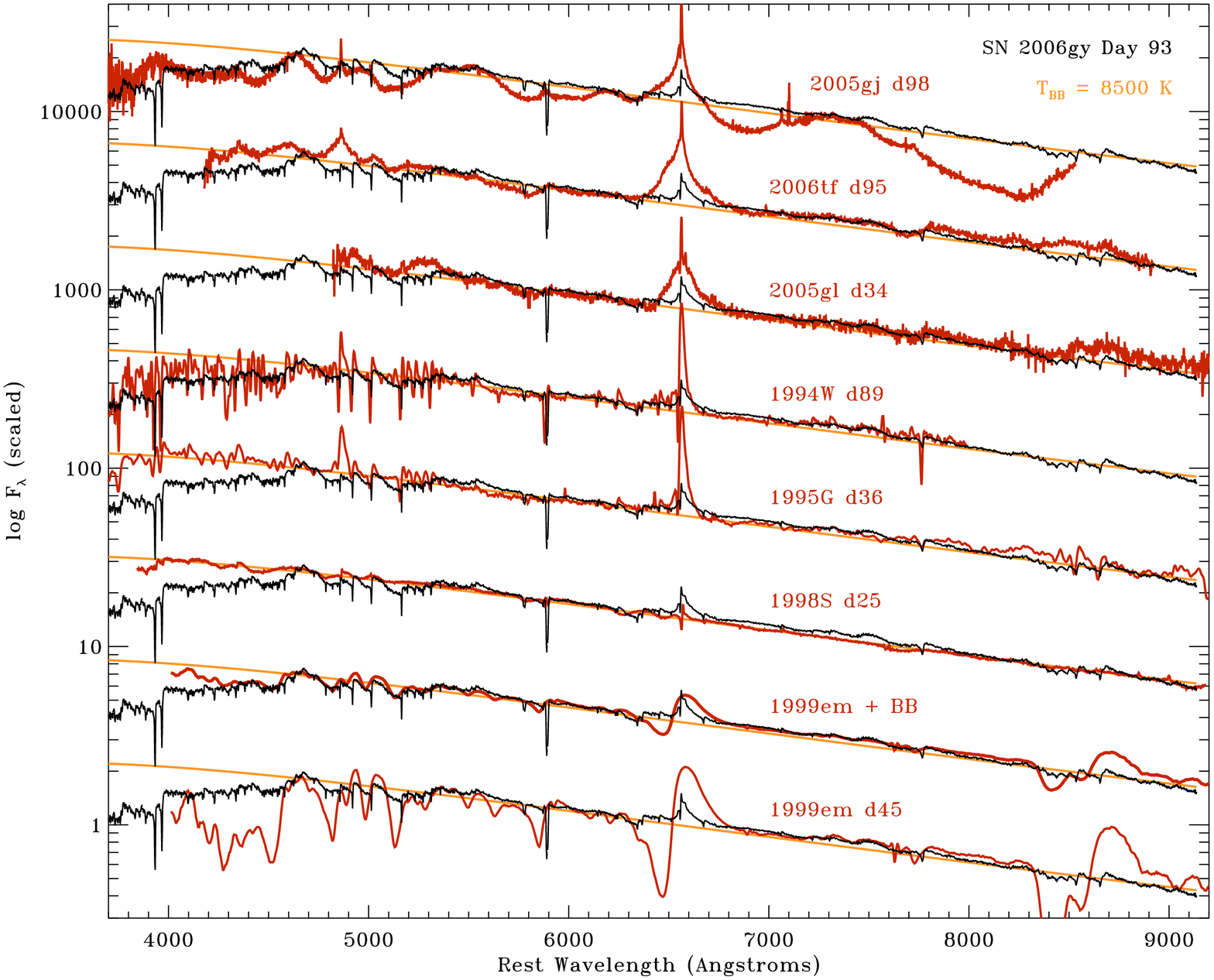}
\caption{The day 93 LRIS spectrum of SN~2006gy compared with spectra of
  several other SNe (red) and with a $T = 8500$~K blackbody (orange). The 
  day 98 spectrum of SN~2005gj is unpublished from our spectral database,
  obtained on 2006 Jan. 01 with DEIMOS at the Keck Observatory.  (In
  order to force a match to the continuum slope of SN~2006gy, this
  spectrum needed to be corrected for $E(B-V) = 0.35$ mag of reddening.
  This is an overcorrection compared to the actual value $E(B-V) = 0.12$
  mag favored by Prieto et al.\ 2007.)  The day 95 spectrum of
  SN~2006tf is from Smith et al.\ (2008a), but again with a high
  reddening correction.  The day 34 spectrum of SN~2005gl is discussed
  in the Appendix.
  The day 89 spectrum of SN~1994W was obtained at Lick Observatory on
  1994 Oct. 11 (see Chugai et al.\ 2004), and was corrected for
  $E(B-V) = 0.17$ mag (Sollerman et al.\ 1998). The day 36 spectrum of
  SN~1995G is from Pastorello et al.\ (2002), and the day 25 spectrum
  of SN~1998S is from Leonard et al.\ (2000).  The day 45 spectrum of
  SN~1999em (bottom) is from Leonard et al.\ (2002) and is
  representative of a normal SN~II-P photosphere.  While this gives a
  poor match to the SN~2006gy spectrum by itself, the general shape of
  the spectrum matches better if we dilute the photospheric SN~II-P
  spectrum with 2/3 of the flux contributed by a $T=8500$~K blackbody.
  In all cases, but especially for SN~1999em, individual line profiles
  do not match because of different velocities in each object.}
\label{fig:comp}
\end{figure*}

\begin{figure*}
\epsscale{0.95}
\plotone{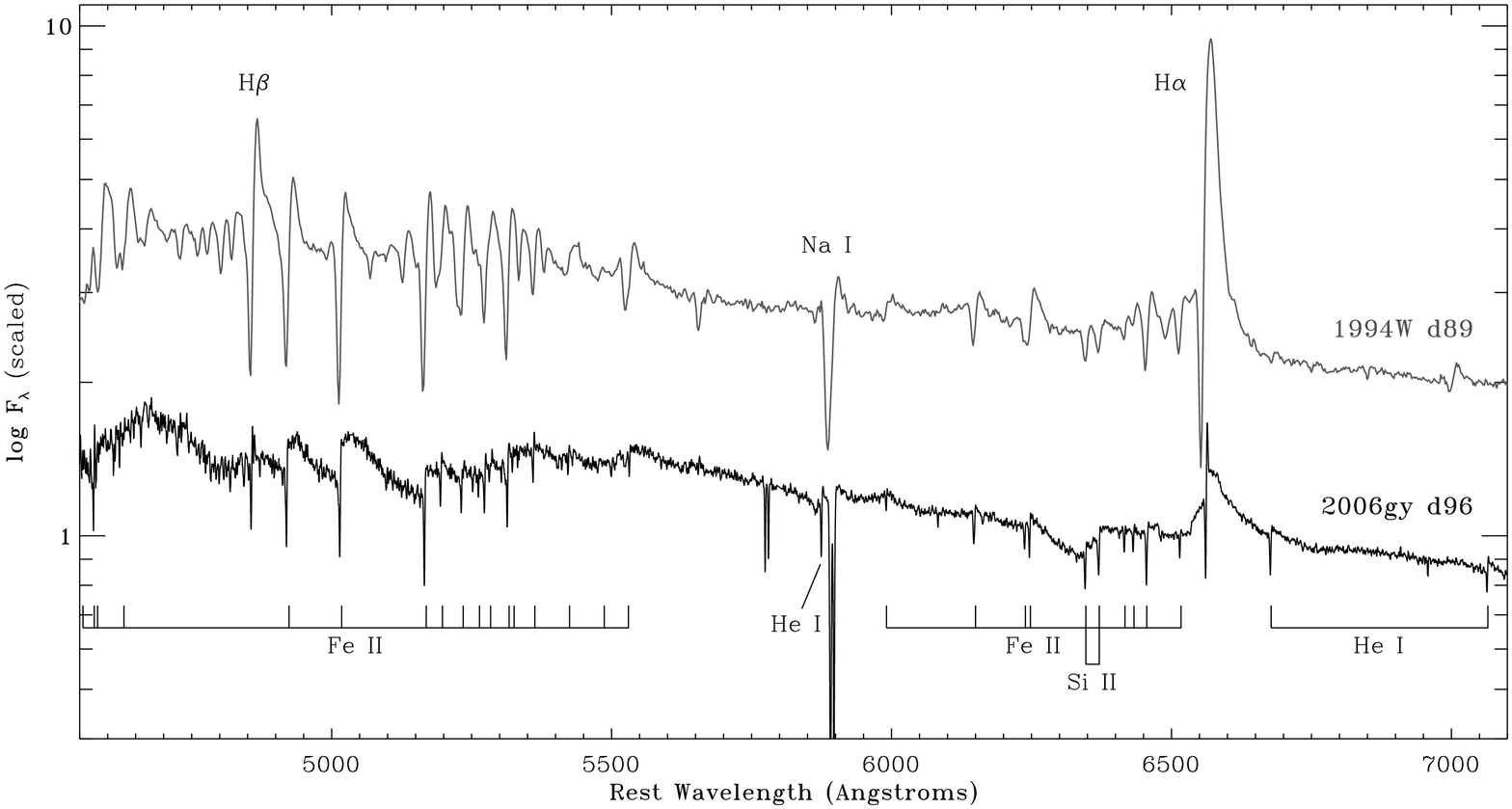}
\caption{The dereddened day 96 high-resolution DEIMOS spectrum of
  SN~2006gy (black) compared to the day 89 spectrum of SN~1994W (gray)
  from Figure~\ref{fig:comp} (see Chugai et al.\ 2004), dereddened
  with $E(B-V) = 0.17$ mag (Sollerman et al.\ 1998).  Several line
  identifications are marked.  While many of the same lines can be
  seen in both SNe, especially in absorption, there are considerable
  differences in the relative emission-line strengths.  Consider the
  different Balmer-line strengths and H$\alpha$/H$\beta$ emission-line 
  intensity ratios, for example.}
\label{fig:lineID}
\end{figure*}

\section{RESULTS}

\subsection{Overall Spectral Morphology and Evolution}

The overall spectral morphology of SN~2006gy can be characterized as a
smooth blue continuum, convolved with a number of striking narrow and
intermediate-width absorption and emission features (see
Figure~\ref{fig:dered}).  Broad absorption features and broad P Cygni
profiles typically seen in SN~II-P atmospheres, with speeds of
$\ga$10$^4$ km s$^{-1}$, are absent in the spectrum during the main
light-curve peak.  Rather, the broadest lines seen in the spectrum
have speeds of $\pm$5000 km s$^{-1}$, where the line wings of H$\alpha$
join the continuum.

During the slow rise to maximum and the peak luminosity phase,
SN~2006gy is characterized by a smooth blue continuum with few
emission features besides the strong Balmer emission lines.  It has
few absorption features other than the interstellar Ca~{\sc ii} and
Na~{\sc i} lines.
Shortly following peak luminosity, SN~2006gy takes on a distinct
spectral morphology that changes little during the decline phase for
$\sim$100 days thereafter.  The spectrum is punctuated by strong
narrow emission and absorption features and intermediate-width
absorption and emission.  These produce interesting line-profile
shapes with smooth red wings, but very sharp drops at $v = 0$ km
s$^{-1}$ and blueward that are different from the smooth P Cygni
profiles normally seen in SNe~II.  The most striking examples of this
are the Fe~{\sc ii} lines near 5000 \AA\ (Figure~\ref{fig:dered}).

The spectrum of SN~2006gy during this decline phase is relatively
unique, but has some overlap with other SNe in various respects.  The
day 93 spectrum of SN~2006gy, which is exemplary of this phase, is
compared with some other SN~II spectra in Figure~\ref{fig:comp}.

{\it SN~2005gj}: Before the discovery of SN~2006gy, SN~2005gj was the
most luminous SN known. It has been suggested to be a Type Ia
explosion interacting with dense H-rich CSM, partly because its
spectrum could be decomposed into a SN~1991T-like spectrum and a
smooth polynomial (Aldering et al.\ 2005; Prieto et al.\ 2007).
SN~2005gj is often grouped together with other luminous SNe~IIn like
SNe~2002ic, 1997cy, and 1999E (Germany et al.\ 2000; Turatto et al.\
2000; Rigon et al.\ 2003; Hamuy et al.\ 2003; Wang et al.\ 2004; Deng
et al.\ 2004; Wood-Vasey et al.\ 2004; Chugai et al.\ 2004; Chugai \&
Yungelson 2004; Kotak et al.\ 2004; Benetti et al.\ 2006; Chugai \&
Chevalier 2007) because their spectra are similar, although the
connection to SNe~Ia has been questioned (Kotak et al.\ 2004; Benetti
et al.\ 2006; Trundle et al.\ 2008).

Agnoletto et al.\ (2009) found an overall remarkable similarity
between the spectrum of SN~2006gy and this class of objects, but our
comparison with SN~2005gj in Figure~\ref{fig:comp} contradicts this
assessment because we find that at a similar phase (93 vs.\ 98 days)
their spectra are very different.  SN~2006gy does not show the broad
absorption features and undulations of the continuum that are seen in
SN~2005gj (and its brethren SNe~2002ic, 1997cy, and 1999E), and its
H$\alpha$ emission is much weaker relative to the continuum.  As we
note below, the similarity improves if one compares later phases of
SN~2006gy to earlier phases of SN~2005gj, but such comparisons at
different phases can be misleading, especially if one draws
connections to the putative type of the underlying SN photosphere.  In
any case, with the high luminosity and long duration of SN~2006gy,
emission from the underlying SN photosphere should make a negligible
contribution to the total spectrum.

{\it SN~2006tf}: SN~2006tf is the second most luminous Type~IIn after
SN~2006gy (Smith et al.\ 2008a).  The similarity between spectra of
SNe~2006gy and 2006tf (days 93 and 95, respectively) is better, as we
noted in Paper~I, although there is still some mismatch in blue
spectral features.  Relative to the continuum, Balmer-line emission is
much stronger in SN~2006tf (note that the intensity scale in
Figure~\ref{fig:comp} is logarithmic).

{\it SN~2005gl}: This SN~IIn deserves special mention, because so far
it is the only SN IIn for which a progenitor star has been detected in
pre-explosion images (Gal-Yam \& Leonard 2009).  It is useful for
comparison because it had a moderate luminosity, typical among SNe~IIn
(Gal-Yam et al.\ 2007).  Early-time spectra were of Type IIn, but by
about 2 months after explosion, the Type~IIn signatures weakened and
the spectrum began to look more like Type II-P.  The day 34 spectrum
that we show in Figure~\ref{fig:comp} resembles that of SN~2006tf,
despite the much lower luminosity of SN~2005gl.  Gal-Yam \& Leonard
(2009) inferred that the progenitor star was a $L/{\rm L}_{\odot}
\approx 10^6$ LBV-like star that, in the year or so before the SN, had
suffered an outburst with $\dot{M} \approx 0.03$ M$_{\odot}$
yr$^{-1}$.  The event was therefore similar to the 1600 AD\ giant
eruption of P Cygni (Smith \& Hartigan 2006). The lower peak
luminosity and relatively quick fading of CSM-interaction signatures
in the spectrum of SN~2005gl are relevant to our study of SN~2006gy,
because the precursor outburst that Gal-Yam \& Leonard estimated,
although extreme compared to normal winds, was tame compared to the
precursor outburst needed for SN~2006gy.

{\it SN~1994W}: Perhaps the best overall similarity to the day 93
spectrum of SN~2006gy in Figure~\ref{fig:comp} is found with the
spectrum of SN~1994W at day 89 (Chugai et al.\ 2004). The shape of the
continuum and the amount of line blanketing at blue wavelengths show
fairly good agreement.  Both SNe share many of the same spectral
features, although the narrow absorption and emission lines are
stronger relative to the continuum in SN~1994W.  A more detailed
comparison to our higher-resolution day 96 DEIMOS spectrum of
SN~2006gy is shown in Figure~\ref{fig:lineID}, where it is clear that
most of the narrow line features are the same in both SNe, despite the
different strengths.  They are substantially narrower in SN~2006gy,
with typical widths of 100--200 km s$^{-1}$ instead of 800--1000 km
s$^{-1}$ in SN~1994W.  Besides Balmer lines, most of these features
are Fe~{\sc ii} (Chugai et al.\ 2004).  Chugai et al.\ interpreted
this emission/absorption spectrum as arising from a combination of an
expanding cold dense shell and dense pre-shock gas expanding at
800--1000 km s$^{-1}$ to explain the narrow absorption features.
Dessart et al.\ (2009), on the other hand, suggested that both
components arise in the recombination photosphere in the slowly
expanding shell of SN 1994W, and that the broad wings arise primarily
from electron scattering.  Given the overall similarity to SN~2006gy,
these aspects will resurface later.

{\it SN~1995G and SN~1998S}: These are two well-studied and
prototypical SNe~IIn with lower luminosity, faster spectral evolution,
and much faster decline rates than those of SN~2006gy (i.e., they
behaved more like SN~2005gl).  Their spectra at 90--100 days after
explosion do not resemble spectra of SN~2006gy at a comparable epoch
because they have already become optically thin by that point (see
Leonard et al.\ 2000; Fassia et al.\ 2001; Pastorello et al.\ 2002).
At earlier times when it was still optically thick, however, the day
36 spectrum of SN~1995G has a strong smooth continuum with narrow
absorption/emission features similar to SN~2006gy, although its Balmer
emission lines are stronger.  Similarly, on day 25 SN~1998S also has a
strong smooth continuum, but in this case the Balmer lines are weaker
than in SN~2006gy.  This gives a qualitative illustration that the
strength of H$\alpha$ is tied to the time evolution of the supernova's
optical depth. For SN~2006gy to maintain high optical depths for so
long is remarkable among SNe~IIn.  We will discuss the H$\alpha$ line
strength in more detail in \S 3.3 and 3.4.  In SNe~1995G and 1998S,
the blue/near-UV continuum is not as heavily affected by line
blanketing as in SN~2006gy or SN~1994W.

{\it SN~1999em}: This is a prototypical SN~II-P spectrum in
mid-plateau.  As noted earlier, one would obviously not expect the
recombination photosphere within a rapidly expanding SN envelope to
match the Type IIn spectrum of SN~2006gy, and indeed the very broad
and deep absorption features in SN~1999em are unlike the narrow lines
in SN~2006gy.  However, we chose to display the spectrum of SN~1999em
for the following reason: if we now {\it dilute} the photospheric
spectrum of SN~1999em with a blackbody (second from bottom tracing in
Figure~\ref{fig:comp}), then we see a striking result.  Ignoring the
discrepant expansion speeds that cause broad lines in SN~1999em and
narrow lines in SN~2006gy, the overall shapes of the two spectra match
surprisingly well.  This is especially true in the blue/near-UV
regions, because this composite SN~1999em+blackbody matches the day 93
spectrum of SN~2006gy better than any SN~IIn in Figure~\ref{fig:comp}.
This may suggest that the spectrum of SN~2006gy is generated by two
components, such as a blackbody from ongoing CSM interaction combined
with a recombination photosphere produced as radiation diffuses slowly
out of an optically thick accelerated H envelope or shell (e.g., Smith
\& McCray 2007).  Alternatively, it may mean that back radiation from
the CSM interaction zone causes a muting of the line features via the
``top-lighting'' mechanism described by Branch et al.\ (2000).
Diluting the observed spectrum in this way also provided the basis for
comparing SN~2005gj to a SN~Ia spectrum (Aldering et al.\ 2006; Prieto
et al.\ 2007).

\begin{figure}
\epsscale{1.05}
\plotone{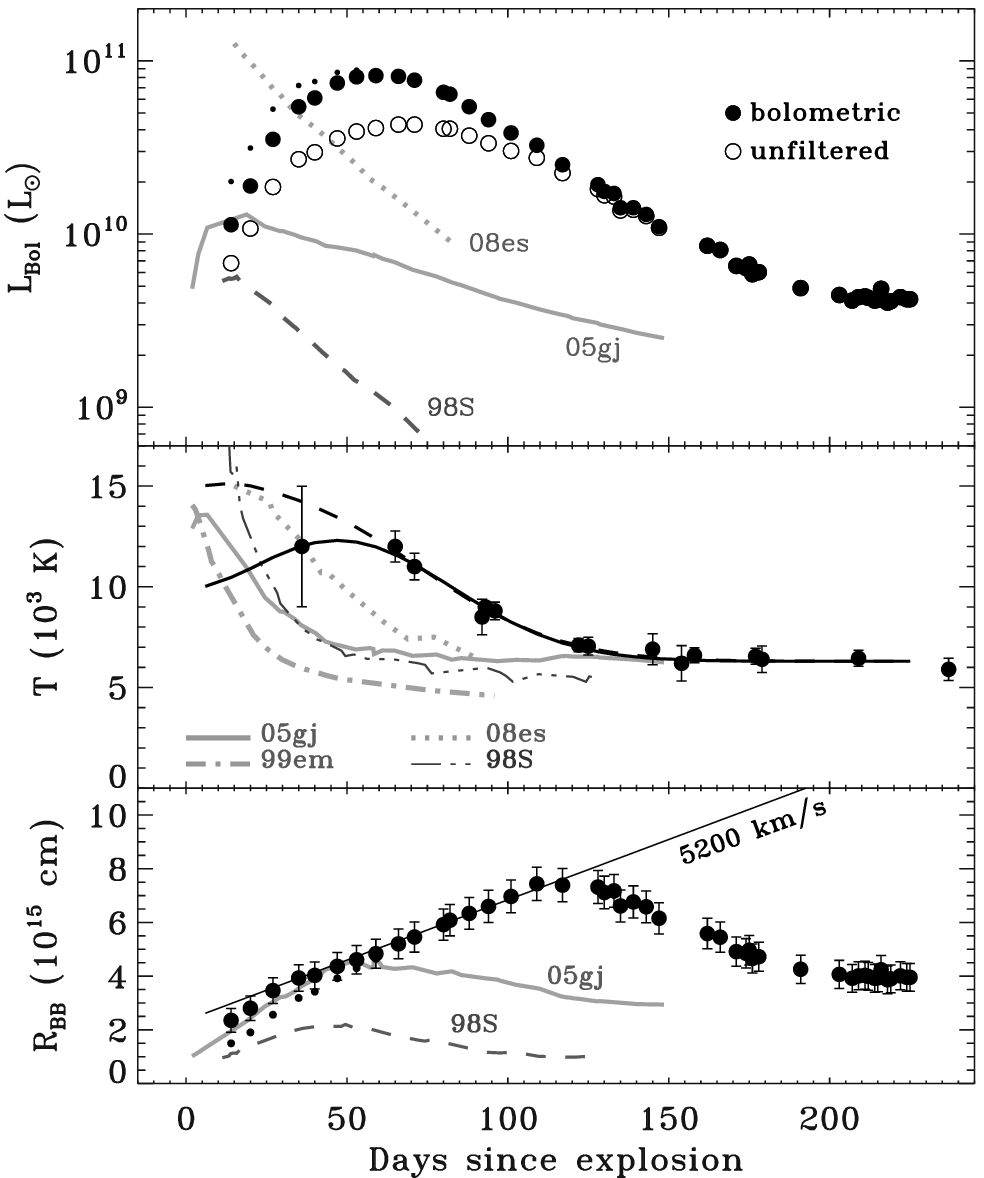}
\caption{The light curve, apparent temperature, and implied blackbody
  photospheric radius of SN~2006gy.  The unfilled points in the light
  curve are derived from unfiltered photometry from the 0.76~m Katzman
  Automatic Imaging Telescope (KAIT, at Lick Observatory) assuming
  zero bolometric correction (from Paper~I).  The large filled and
  small filled circles in the light curve correspond to two different
  bolometric corrections that were applied, derived from the two
  temperature curves (solid and dashed, respectively) in the middle
  panel under two different assumptions about the early times (see
  text).  Filled points plotted in the middle panel are apparent
  temperatures derived from observed continuum shapes in our
  dereddened spectra (Fig.~\ref{fig:dered}; Table 1).  The bottom
  panel shows the blackbody radius derived from the
  bolometric-corrected luminosity and the apparent temperature.
  Again, the large and small filled circles correspond to two
  different assumptions about the early-time temperature evolution.
  For comparison ({\it gray}), we also show bolometric luminosities
  and blackbody temperatures for SN~2008es (Miller et al.\ 2009a;
  Gezari et al.\ 2009), SN~1998S (Fassia et al.\ 2000), SN~2005gj
  (Prieto et al.\ 2007), and the prototypical Type II-P SN~1999em
  (Leonard et al.\ 2002).  In the bottom panel, 5200 km s$^{-1}$
  corresponds to the full width at half-maximum intensity (FWHM) of
  the broad Gaussian component fit to the H$\alpha$ profiles described
  later.  The inferred blackbody radii for SN~1998S and SN~2005gj are
  also shown.}
\label{fig:lc}
\end{figure}

\subsection{Continuum Evolution}

The dereddened visual-wavelength continuum shape can be approximated
adequately with a blackbody function, with the caveat that there is
substantial line-blanketing absorption at blue wavelengths,
particularly severe below $\sim$4500 \AA.  No single temperature could
account for the spectral shape both above and below 4500 \AA, so in
Figure~\ref{fig:dered} we show blackbody curves ({\it orange}) which
approximate the $\lambda > 4500$ \AA\ continuum shape quite well,
assuming that the decrease at short wavelengths is caused by
additional absorption of the emitted blackbody spectrum.  (The
blue-wavelength flux at late times is quite uncertain and sensitive to
the background subtraction.)  This drop in flux at short wavelengths
is common in SN~II-P atmospheres, as noted above, and in some SNe~IIn.
A few representative temperatures for the continuum of SN~2006gy are
shown in Figure~\ref{fig:dered}, but all are listed in Table 1.

The evolution of the apparent blackbody temperature is plotted in
Figure~\ref{fig:lc} (middle panel), where the solid black line is a
smooth functional approximation to the time-dependent temperature,
since our spectra irregularly and sparsely sample the temporal
evolution.  The temperature at early times (before day 50) is poorly
constrained by our data because of the substantial reddening
correction, and because of calibration issues with our day 36 spectrum
as noted in \S 2.  The day 36 estimate of the temperature has large
uncertainty, so we consider two different cases for the early-time
temperature evolution: a decline to lower temperatures around 10$^4$~K
(solid curve) and a temporary rise to higher peak temperatures near
15,000~K (dashed).  The lower temperature gives a better approximation
of the continuum shape if we adopt the same reddening for all epochs
(dashed day 36 spectrum in Figure~\ref{fig:dered}).  However, as shown
in Figure~\ref{fig:dered}, one can approximate the continuum shape
somewhat better with a hotter 15,000~K blackbody if the day 36
spectrum is corrected for a larger reddening than the other epochs.
This interpretation seems plausible given the high optical depths we
expect in the inner CSM shell, as discussed later.  We do not expect
that the temperature was much above 15,000 K, since the day 36
spectrum shows no sign of the strong WR-like N~{\sc iii} and He~{\sc
  ii} features seen in the early phases of SN~1998S with temperatures
above 20,000~K (Leonard et al.\ 2000).

These two treatments of the temperature yield somewhat different
bolometric corrections and emitting radii at early times
(Fig.~\ref{fig:lc}), but have little impact on our overall
interpretation.  The solid curve would suggest a temperature evolution
different from that of other SNe~IIn, whereas the dashed curve with
higher temperatures would make the temperature of SN 2006gy decline
qualitatively similar to other SNe~IIn.  In either case, SN~2006gy
takes much longer to cool, which is no doubt related to the
extraordinary longevity of its peak luminosity phase.  By day 150, the
temperature settles to a floor at $\sim$6500~K and remains nearly
constant thereafter.  The fact that this temperature agrees with the
same late-time value inferred for less-reddened SNe~IIn (like
SNe~2005gj and 1998S in Fig.~\ref{fig:lc}) could be taken to indicate
that our assumed reddening value of $E(B-V) = 0.72$ mag is roughly
correct.  This temperature floor appears to be common among SNe~IIn,
whereas SNe~II-P continue their decline to somewhat lower temperatures
during their plateau phase.

These smooth curves approximating the temperature change of SN 2006gy
are used to derive an approximate bolometric correction for the
observed light curve in Figure~\ref{fig:lc} (top panel).  The large
filled circles correspond to the lower temperatures at early times
(solid curve in temperature), whereas the small circles show what the
corrected luminosity would be with the higher assumed temperature
(dashed curve), although not taking into account the larger extinction
correction implied by the higher temperature (see
Fig.~\ref{fig:dered}).
Warmer temperatures before day 110 lead to an increase in our estimate
of the bolometric luminosity (solid circles) and a slight shift in the
time of peak luminosity from 70 days to about 60 days after the
nominal explosion date.  This bolometric correction affects the
measured total radiated energy as well: in Paper~I we found that the
radiated energy integrated over the first $\sim$200 days was $1.6
\times 10^{51}$ erg.  With the corrected $L_{\rm Bol}$ light curve in
Figure~\ref{fig:lc}, this estimate rises to roughly (2.3--2.5) $\times
10^{51}$ erg (the uncertainty is due to the uncertain temperature
evolution at early times).

With these bolometric luminosity and temperature curves, we also
derive the effective blackbody radius, $R_{\rm BB}^2 = L / (4 \pi
\sigma T^4)$, plotted in the bottom panel of Figure~\ref{fig:lc}.
Again, the large solid points with error bars are for the lower
temperatures at early times, and the small points show the derived
radius for the higher temperatures.  Differences between these two are
minor, and do not much affect our discussion.  The effective emitting
radius of SN~2006gy behaves in a manner qualitatively similar to those
of other SNe~IIn in that $R_{\rm BB}$ increases steadily for a time,
reaching a peak and then decreasing again.  For SN~2006gy, this
turnover occurred much later than in any other SN --- $R_{\rm BB}$
turned over around day 115, whereas it occurred much earlier at day
$\sim$50 for both SNe~1998S and 2005gj (Fassia et al.\ 2000; Prieto et
al.\ 2007), and at day 70--80 in SN~2006tf (Smith et al.\ 2008a).  In
all cases, the turnover of $R_{\rm BB}$ was well after the time of the
supernova's peak luminosity.

Presumably this peak represents a critical transition point when the
shell starts to become optically thin.  In the case of SNe~II-P, this
occurs as the H-recombination photosphere recedes backward through the
geometrically thick expanding H envelope of the star.  This marks a
clear division in radius between the inner optically thick ionized
ejecta and the outer recombined transparent ejecta.  In SNe~IIn,
however, one expects the post-shock shell to form a geometrically very
thin layer (a ``cold dense shell,'' or CDS; see Chugai et al.\ 2004),
because the post-shock layer collapses due to radiative cooling.  In
that case, the turnover and subsequent decrease of $R_{\rm BB}$ is
misleading, because $R_{\rm BB}$ decreases more than the expected
shell thickness, which should be much less than 10\% of $R_{\rm
  shell}$.  Also, apparent velocities measured in the H$\alpha$ line
do not decrease during this time.

Motivated by similar observations of the luminous SN~2006tf, we (Smith
et al.\ 2008a) explained this behavior in SNe~IIn as follows: the
derived value of $R_{\rm BB}$ is not the true radius of the emitting
surface in the cold dense shell, but rather, $R_{\rm BB}^2 = \zeta
R_{\rm shell}^2$.  The true physical radius of the shell, $R_{\rm
  shell}$ continues to increase with time, but $\zeta$ is a dilution
factor representing the fractional geometric covering area of the
shell ($\zeta \le 1$, decreasing with time).  The post-shock shell is
able to continue producing continuum radiation with a constant
effective temperature of 6000--6500 K, despite its reduced effective
opacity, because it is clumpy; small clumps remain ionized and
optically thick longer, whereas less dense interclump regions
gradually recombine and become transparent.  One expects the CDS to be
severely clumped because of Rayleigh-Taylor instabilities at the
contact discontinuity.  Limb-brightening effects may also be important
in determining $\zeta$.

The rate at which $R_{\rm BB}$ increases with time during early phases
before this transition occurs (between days 40 and 120) seems to
roughly match an expansion speed of $\sim$5,200 km s$^{-1}$ (solid
line in Fig.~\ref{fig:lc}).  This speed is taken from the broadest
Gaussian matched to the H$\alpha$ emission profiles, which seem to
imply constant expansion speeds at all times, as we discuss below in
\S 3.3.  H$\alpha$ continues to indicate roughly the same expansion
speeds after the turnover in $R_{\rm BB}$ at day 115, when one expects
that high optical depth effects like electron scattering wings will be
less influential.  The expansion speed of the photosphere seemed to be
somewhat faster before day 25, but unfortunately our data do not
provide good constraints because of uncertainties associated with the
apparent temperature at early times.  The radius around day 25 of $\la
3 \times 10^{15}$ cm or $\la$200~AU roughly matches the expected
initial radius of the highly opaque CSM shell inferred by Smith \&
McCray (2007), who claimed that it may have taken a few weeks for the
SN to traverse this shell and become luminous.

\begin{figure}
\epsscale{0.97}
\plotone{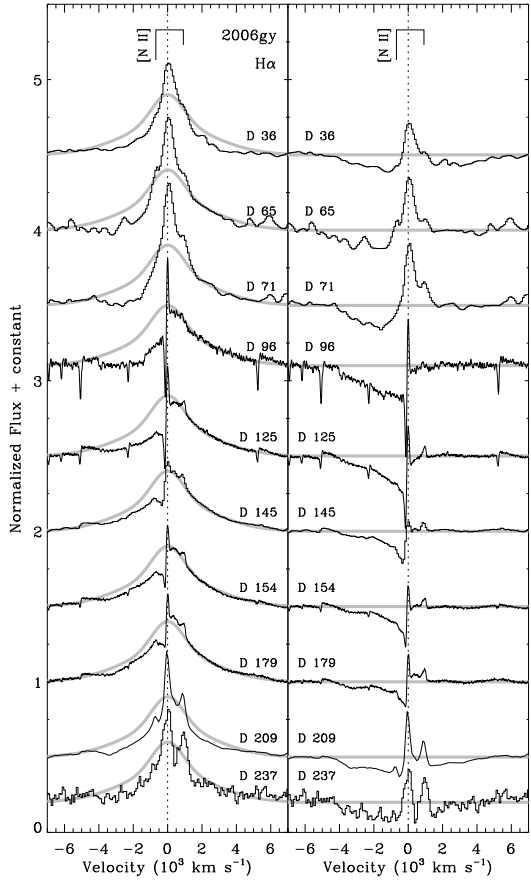}
\caption{H$\alpha$ line profiles in SN~2006gy at several epochs (not
  including redundant spectra with lower resolution).  The left side
  shows the observed profile normalized to the continuum level,
  overplotted with a symmetric Gaussian curve in gray.  This is a sum
  of two Gaussians with a FWHM of 5200 and 1800 km s$^{-1}$, each with
  the same peak intensity.  The expected positions of [N~{\sc ii}]
  $\lambda\lambda$6548, 6583 are shown.  The right side shows the same
  H$\alpha$ profiles with the symmetric Gaussian subtracted to
  emphasize the blueshifted broad absorption.  The first three spectra
  at days 36, 65, and 71 have lower spectral resolution.}
\label{fig:haAll}
\end{figure}

\begin{figure}
\epsscale{0.99}
\plotone{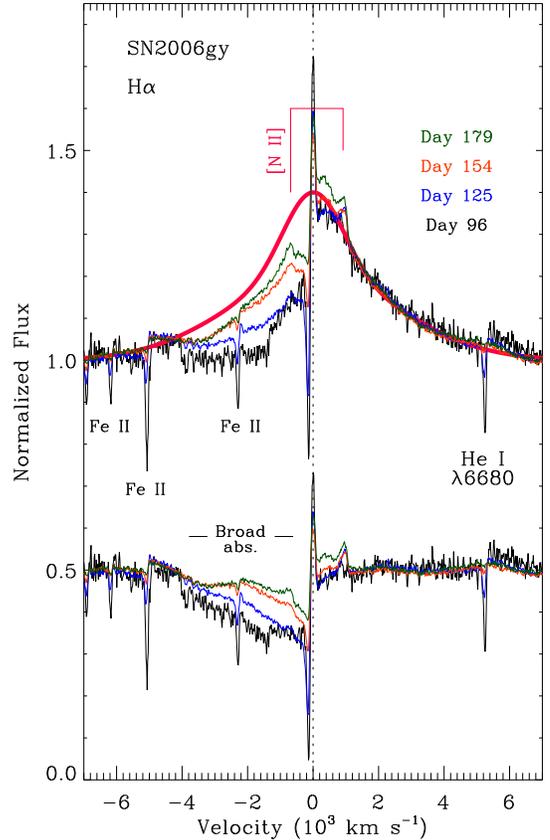}
\caption{The broad component of the H$\alpha$ line profile in
  SN~2006gy from the four DEIMOS spectra obtained at high resolution,
  normalized to the same continuum level and overplotted.  The blue
  wing of H$\alpha$ undergoes strong and systematic
  variations, with the absorption weakening with time, while the red
  wing remains remarkably constant.  The expected positions of [N~{\sc
    ii}] $\lambda\lambda$6548, 6583 are shown.  The thick magenta curve
  is the sum of two symmetric Gaussians with a FWHM of 5200 and 1800 km
  s$^{-1}$, as in Fig.~8.  The lower set of tracings are the same as
  the upper profiles, but with this Gaussian subtracted, as in
  Fig.~8.}
\label{fig:haNorm}
\end{figure}

\subsection{The Broad H$\alpha$ Line Profile}

The evolution of the broad to intermediate-width component of
H$\alpha$ is a key diagnostic of the CSM interaction in SNe~IIn, and
will be discussed in some detail in the following sections.  We show a
time series of the H$\alpha$ profile of SN 2006gy in
Figure~\ref{fig:haAll}, and we display the line profiles at several
epochs obtained with high resolution in Figure~\ref{fig:haNorm}.

Caveats to keep in mind for the discussion below are the following.
Our first three spectra on days 36, 65, and 71 were obtained
with the Kast spectrograph at Lick Observatory, and have lower
spectral resolution than the Keck/DEIMOS and LRIS data.  This alters
the appearance of the narrow emission/absorption components and the
transition between the narrow and broad components on those dates in
Figure~\ref{fig:haAll}.  The red side of the H$\alpha$ line is
affected by [N~{\sc ii}] $\lambda$6583 emission to varying degrees,
and the blue side by [N~{\sc ii}] $\lambda$6548 to a lesser degree;
the positions of these lines relative to the H$\alpha$ systemic
velocity are marked in Figures~\ref{fig:haAll} and \ref{fig:haNorm}.
The H$\alpha$ profile is also affected by weaker broad and narrow
lines of He~{\sc i} and Fe~{\sc ii}, as noted in Paper~I, whereas the
blue wing of the line is depressed by varying amounts of broad
blueshifted H$\alpha$ absorption.  For all profiles in
Figure~\ref{fig:haAll}, we have normalized to the smooth continuum
level (see Fig.~\ref{fig:dered}) and scaled the emission-line strength
in order to compare profile shapes.  We do not plot the H$\alpha$
profiles on days 92, 93, 122, 127, 153, 158, and 177, because they are
redundant for our purposes, superseded by higher-resolution DEIMOS
data obtained within a few days.

To provide a fiducial profile for comparison, we plot each H$\alpha$
profile on top of a symmetric Gaussian, shown in gray in
Figure~\ref{fig:haAll} and in magenta in Figure~\ref{fig:haNorm}.
This symmetric profile is the sum of two Gaussians with equal
intensity and FWHM values of 5200 km s$^{-1}$ and 1800 km s$^{-1}$.
It is not necessarily the best fit to the profile on any specific
date, but it gives an adequate match to the red wing for most epochs
except as noted below.  In addition, for the right-hand panel in
Figure~\ref{fig:haAll} and for the lower set of tracings in
Figure~\ref{fig:haNorm}, we show the residual profile after
subtracting away this symmetric Gaussian in order to emphasize changes
in the broad blueshifted H$\alpha$ absorption.

\begin{figure} 
\epsscale{0.99}
\plotone{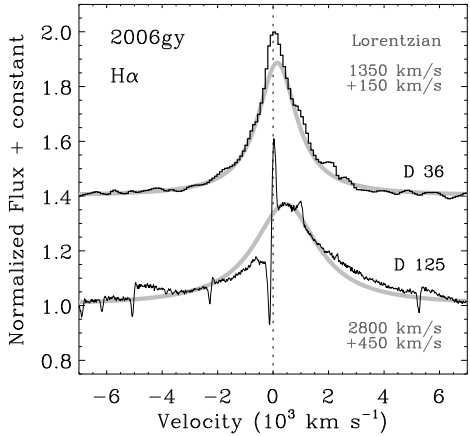}
\caption{The H$\alpha$ profiles on days 36 and 125, compared to
  Lorentzian profiles.  The day 36 spectrum can be adequately
  represented by a single Lorentzian with FWHM = 1350 km s$^{-1}$ and
  with its centroid redshifted by 150 km s$^{-1}$, ignoring some
  unresolved narrow emission.  A Lorentzian with FWHM = 2800 km s$^{-1}$
  and with its centroid redshifted by 450 km s$^{-1}$ does not give a
  very good representation of the day 125 line profile, or the
  H$\alpha$ profile at any other epoch after the time of peak
  luminosity.}
\label{fig:haLor}
\end{figure}

\subsubsection{The Pre-Maximum H$\alpha$ Profile}

Our earliest spectrum --- the only one obtained while SN~2006gy was
still on its slow rise to maximum luminosity --- was obtained on day
36.  This is an unfortunate consequence of the fact that the SN was
not discovered until about 29 days past our adopted explosion date.
(Upper limits and photometry before that time were extracted from
pre-discovery data.)  Also, SN~2006gy was initially thought to be an
active galactic nucleus; its peculiar SN nature was not realized until
relatively late (see Paper~I and references therein).

As discussed below, the broad Gaussian curve in Figure~\ref{fig:haAll}
gives an adequate match to the red wing of H$\alpha$ for most epochs,
with a clear exception being the first spectrum on day 36.  The
broadest component on day 36 is substantially {\it narrower} than the
Gaussian curve; consequently, subtraction residuals show a broad
deficit in the redshifted wings.

Instead, the broad H$\alpha$ wings on day 36 are better matched by a
single Lorentzian profile, with FWHM = 1350 km s$^{-1}$ and with its
centroid redshifted by $+$150 km s$^{-1}$ (Fig.~\ref{fig:haLor}).
This is not true for any epoch after day 90, when a single Lorentzian
that fits the broad line wings and core would significantly
underestimate the wings at $\pm$1000--3000 km s$^{-1}$, and would
require a net redshift of $+$450 km s$^{-1}$ (an example of such a
profile plotted with the day 125 spectrum is shown in
Figure~\ref{fig:haLor}).  In addition to the broad Lorentzian, the
line has an unresolved narrow Gaussian component that contributes
$\sim$7\% of the total H$\alpha$ luminosity.

A similar result was found for the H$\alpha$ profile in early spectra
of SN~1998S, where Leonard et al.\ (2000) showed that on day 5 it was
well fit by a broad Lorentzian plus an unresolved component that
contributed 20--25\% of the line luminosity.  Chugai (2001) proposed
that this profile arises from an intrinsically narrow H$\alpha$ line
that is broadened by collisions with thermal electrons in an opaque
circumstellar shell outside the photosphere.  The Lorentzian wings
arise from multiple electron scattering, whereas the unresolved narrow
component is direct emission.  A similar model may explain the
early-time H$\alpha$ emission from SN~2006gy.  In this interpretation,
the net $+$150 km s$^{-1}$ redshift of the Lorentzian centroid
corresponds to the expansion speed of the CSM at these inner radii
(see Chugai 2001).  A rough estimate of the scattering envelope's
optical depth outside the photosphere, $\tau$, is given by $U = (1
-e^{-\tau}) / \tau$, where $U$ is the ratio of the unscattered narrow
component to the total line luminosity.  For SN~2006gy on day 36 we
measure $U \approx 0.07$, yielding $\tau \approx 15$.  The pre-SN
envelope is evidently highly opaque.

The FWHM of this Lorentzian component on day 36 is only 1350 km
s$^{-1}$, which is much slower than the necessary expansion speed of
$\sim$5000 km s$^{-1}$ corresponding to $d(R_{\rm BB})/dt$ in
Figure~\ref{fig:lc}.  If anything, the true expansion speed of the
photosphere should have been {\it even faster} before day 36.  Thus,
at this early phase, the line wings cannot be caused by electron
scattering within the rapidly expanding CDS itself, as suggested for
SN~1994W by Dessart et al.\ (2009), although this effect may play a
role at later epochs (see below).  Since no broader components are
seen at early times, we must conclude that the H$\alpha$ profile on
day 36 does not trace the kinematics of the SN.  Rather, it seems
likely that the opaque photosphere occurs {\it outside} the forward
shock, as a radiative precursor ionizes the relatively slow and
extremely dense pre-shock CSM.  This opaque ionized region hides the
underlying SN kinematics from our view, so that the broad line
profiles at early times are due entirely to Thomson scattering of
narrow lines formed in the CSM, whereas the rising luminosity and
increase in $R_{\rm BB}$ at early times is due entirely to a rapidly
expanding ionization front.  The dereddened H$\alpha$/H$\beta$ ratio
of $\sim$2.9 on day 36 (Table 1) seems consistent with this
interpretation.  This radiation cannot escape directly from the
post-shock zone, but must diffuse out through the opaque envelope, as
suggested by Smith \& McCray (2007) to explain the slow rise to
maximum of SN~2006gy.

Altogether, the spectrum of SN 2006gy in the time period when it was
still slowly rising to maximum luminosity can be characterized by a
blackbody photosphere expanding within an opaque electron-scattering
circumstellar shell.  This was also the case for SN~1998S, but only at
the very earliest times up to day 5. The much longer duration of this
phase in SN~2006gy is yet another testament to the extraordinarily
dense and extended pre-SN envelope.  It is unfortunate that additional
spectra were not obtained during this early phase; they would have
allowed us to constrain the density structure in the inner parts of
this circumstellar shell in more detail.

\subsubsection{Nearly Constant-Velocity Line Wings}

A striking conclusion from examination of Figures~\ref{fig:haAll} and
\ref{fig:haNorm} is that after the time of peak luminosity, the profile
of the broad emission component of H$\alpha$ is remarkably constant.  Most
importantly, the speed does not decrease substantially as the SN emits
most of its radiation.

To be more precise, the line width does change somewhat with time --
in fact, the line width {\it increases} with time after peak
luminosity (Fig.~\ref{fig:haNorm}) -- but this is primarily because
varying amounts of P Cygni absorption alter the blue wing, as
discussed below.  Examining the red wing of H$\alpha$ exclusively,
however, we see {\it no significant changes in the underlying emission
  profile shape from the time of peak luminosity onward}.  This is
especially clear in the residuals after the same symmetric Gaussian
has been subtracted from each profile (Figs.~\ref{fig:haAll}[right]
and \ref{fig:haNorm}).  Except for a small bump around He~{\sc i}
$\lambda$6678 and narrow emission from [N~{\sc ii}], the red-wing
subtraction residuals are essentially flat for all epochs after
day~36.  In particular, we see no evidence for a change in expansion
speed before, during, or after the time when $R_{\rm BB}$ turns over
(Fig.~\ref{fig:lc}) in transition to the optically thin phase.

Observations of other SNe~IIn tend to show declining FWHM for the
post-shock gas during high-luminosity phases, as for SN~1988Z in its
first year (Turatto et al.\ 1993; Chugai \& Danziger 1994; Aretxaga et
al.\ 1999), whereas models predict a sharp decline in the CDS velocity
around the time of peak luminosity (Chugai 2001; Chugai \& Danziger
2003; Chugai et al.\ 2004).  A nearly constant expansion speed for the
CDS is a product of these models at later times, however, in cases
where most of the deceleration has already occurred at early times
near peak luminosity when the blast wave overtook a dense shell, as
suggested for SN~1994W and SN~1995G (Chugai \& Danziger 2003; Chugai
et al.\ 2004).  Subsequently, a constant shell speed is seen as the
forward shock takes on self-similar expansion through a $\rho \propto
r^{-2}$ wind.  At early times near peak luminosity, however, models
predict that the CDS emitting the intermediate-width H$\alpha$ line
decelerates as the post-shock cooling zone is mass loaded by the CSM
gas swept up by the forward shock.  As we have argued above, this
early deceleration phase was hidden from our view by the opaque CSM
that surrounded SN~2006gy.

Thus, in principle, a constant expansion speed of the post-shock shell
in SNe~IIn is understandable, but the case of SN~2006gy is problematic
because of the energy budget.  SN~2006gy remained at high luminosity
for an astonishingly long time, emitting more than 10$^{51}$ erg of
visual light while its expansion speed remained roughly constant.  One
expects 10$^{51}$ erg to be a substantial fraction of the total
available kinetic energy of the explosion, so unless the ejecta of
SN~2006gy are extremely massive and energetic, the lack of
deceleration as this energy escapes by radiation is puzzling, as we
noted in Paper~I.  Furthermore, after day $\sim$120 when $R_{\rm BB}$
retreats and the shell begins to become optically thin, the luminosity
is still much higher than one can expect from the level of CSM
interaction inferred from the H$\alpha$ or X-ray luminosities, as
discussed below.  A possible resolution to this paradox, as suggested
by Smith \& McCray (2007), is if the expected deceleration occurs at
earlier opaque phases when the radiation cannot escape, converting
shock energy into thermal energy, and emitting after a delay because
that radiation must diffuse out of the opaque shell.

While the velocities up to 5200 km s$^{-1}$ that we infer from the
broad H$\alpha$ component are consistent with the rate of increase of
$R_{\rm BB}$, we cannot rule out the possibility that other factors affect
the line profile.  Namely, Dessart et al.\ (2009) suggested that in
the case of SN~1994W, which appears to have similar line widths and a
similar overall spectral morphology to SN~2006gy, the line width was
generated largely by incoherent electron scattering within the
H$\alpha$-emitting shell itself. SN~2006gy is a good candidate for
investigating high optical depth effects, and further hydrodynamic and
radiation-transfer models are encouraged.  Since the photosphere radius
must be increasing at speeds near 5000 km s$^{-1}$ that are comparable to
the expected electron-scattering wings, however, one expects that the
kinematic and electron-scattering wings are intermixed and
hydrodynamics cannot be treated separately from radiation transfer.

Our analysis of the H$\alpha$ line profiles is different from that of
Agnoletto et al.\ (2009), who find a broad Gaussian component in
H$\alpha$ with FWHM = 9000 km s$^{-1}$.  Their conclusion was based on
a Gaussian decomposition of the H$\alpha$ profile seen in a
low-resolution, late-time (day 173) spectrum.  Our spectra show that
at such late times, the spectral regions on either side of H$\alpha$
are affected by additional broad absorption features unrelated to
H$\alpha$, and these can mimic the appearance of broader line wings.
On the red wing of H$\alpha$, there is an absorption feature below the
nominal continuum level at 6700--7000 \AA, and the blue wing of
H$\alpha$ is affected by a complex blend of broad Si~{\sc ii}
absorption and several Fe~{\sc ii} emission features.  Choosing the
trough of this absorption as the continuum gives an underlying
continuum level and slope that are inconsistent with the rest of the
spectrum across a broader wavelength range.  This red absorption
feature is seen even more clearly in the cases of other luminous
SNe~IIn like SN~2005gj (Fig.~\ref{fig:comp}).  Thus, our data show no
evidence for an H$\alpha$ emission component broader than 5200 km
s$^{-1}$ that might be attributed to fast underlying SN ejecta.  SN
ejecta moving at 9000 km s$^{-1}$ would have already reached and
overtaken the reverse shock (with $R \le R_{\rm BB}$; Fig.~\ref{fig:lc})
by day $\sim$70 when the post-shock region was still opaque.  All
ejecta reaching the reverse shock after that point must be slower
unless the ejecta and CSM are severely asymmetric.

\subsubsection{Broad Blueshifted Absorption}

SN~2006gy is quite extraordinary among SNe~IIn in that its
intermediate-width H$\alpha$ line from the post-shock shell also
exhibits significant blueshifted absorption.  Excluding narrow
absorption from pre-shock CSM gas, most SNe~IIn, by contrast, show
more symmetric H$\alpha$ wings, with some key examples being SN~1988Z
(Turatto et al.\ 1993), SN~1997eg (Salamanca et al.\ 2002; Hoffman et
al.\ 2008), SN~1997cy (Turatto et al.\ 2000), SN~1999E (Rigon et al.\
2003), and SN~2006tf (Smith et al.\ 2008a).  The broad P Cygni
absorption in SN~2006gy is, however, reminiscent of weaker broad
absorption at similar speeds in SN~1994W around day 50 (Chugai et al.\
2004; Dessart et al.\ 2009) and SN~1998S at 2--4 weeks (Fassia et al.\
2001).  In Paper~I we noted that this broad blueshifted absorption
extends out to roughly $-$4000 km s$^{-1}$, seen most clearly in the
day 96 profile when a sharp blue edge is apparent.  Here we find that
some broad P Cygni absorption is present at most epochs.

In Figures~\ref{fig:haAll} and \ref{fig:haNorm}, the nature of this
flux deficit as absorption (as opposed to line asymmetry) is
particularly clear when we subtract away a symmetric Gaussian profile.
Interestingly, these Gaussian-subtracted profiles look similar to the
H$\alpha$ profile of SN~1998S observed on day 17.7 (Fassia et al.\
2001), which showed broad absorption out to $-$6000 km s$^{-1}$ and a
narrow P~Cygni component, but no broad emission.  In SN~2006gy,
residuals show consistent blueshifted absorption at 0 to $-$4000 km
s$^{-1}$.  Admittedly, the strength of this P Cygni absorption on day
36 is unclear, as we have argued above that the line can also be fit
well with a narrower Lorentzian profile that has a redshifted centroid
as a natural consequence of electron scattering in an expanding wind.
The remaining epochs, however, cannot be explained in this way because
their red-wing subtraction residuals are flat.

After day 36, the relative strength of this broad P Cygni absorption
changes with time compared with the emission profile.  It seems to
increase during and immediately after peak luminosity, being strong
from days 71 to 125.  The absorption is strongest in the day 96
spectrum, and weakens thereafter as the SN fades
(Fig.~\ref{fig:haNorm}).  Because the broad blueshifted absorption
steals a substantial fraction of the total line flux at some epochs,
it is relevant to our interpretation of the H$\alpha$ equivalent width
and line luminosity (\S 3.4).

Although the {\it strength} of the broad blueshifted absorption
changes with time, the total {\it range} of velocities out to roughly
$-$4000 km s$^{-1}$ remains constant from the time of peak luminosity
until our last spectrum over 150 days later.  This constant speed seen
in absorption is consistent with the unchanging speeds deduced from
the red wing of H$\alpha$, supporting the notion that the CDS and
blast wave of SN~2006gy continue to expand at constant speed.  The
maximum speed of absorbing material along our sight-line is 4000 km
s$^{-1}$, somewhat less than the FWHM of the broadest Gaussian
emission component and less than the nominal expansion speed indicated
by the growth of $R_{\rm BB}$ in Figure~\ref{fig:lc}.  The
significance of this difference is not immediately clear, and may
require detailed radiative transfer models of the line profiles.  For
example, it is possible that the true expansion speed is 4000 km
s$^{-1}$, while the broader emission wings out to $\pm$6000 km
s$^{-1}$ are due to electron scattering.

Interestingly, the broad blueshifted absorption never dips below the
continuum level.  At its deepest point on day 96, the blueshifted
absorption floor is flat at roughly the continuum level.  This may
suggest primarily self-absorption of the H$\alpha$ line wings and not
necessarily absorption of the continuum.

\begin{figure}
\epsscale{0.99}
\plotone{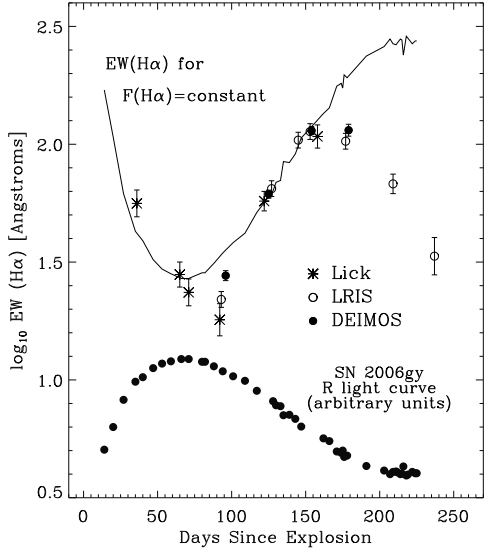}
\caption{Observed H$\alpha$ equivalent width in spectra of SN~2006gy.
  The light curve is shown for reference at the bottom, and the solid
  curve shows what the H$\alpha$ equivalent width would be if the
  H$\alpha$ emission-line flux were constant and all that changed were
  the underlying continuum luminosity.  The three plotting symbols
  refer to measurements with three different instruments (see Table
  1).}
\label{fig:haEWgy}
\end{figure}

\begin{figure}
\epsscale{0.99}
\plotone{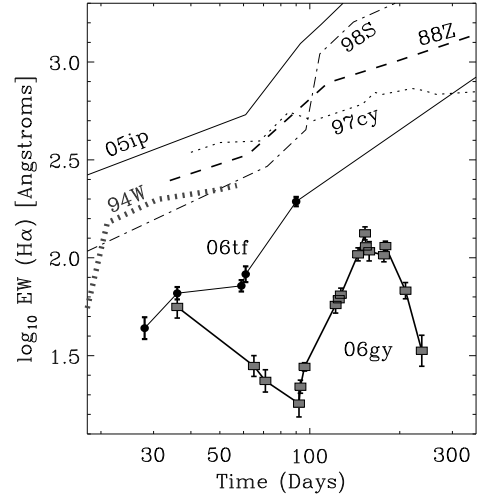}
\caption{Observed H$\alpha$ equivalent width in spectra of SN~2006gy
  from Figure~\ref{fig:haEWgy} compared to several other SNe (compiled
  by Smith et al.\ 2008a, 2009a).}
\label{fig:haEW}
\end{figure}

\subsection{The Total H$\alpha$ Equivalent Width and Luminosity}

\begin{figure}
\epsscale{0.99}
\plotone{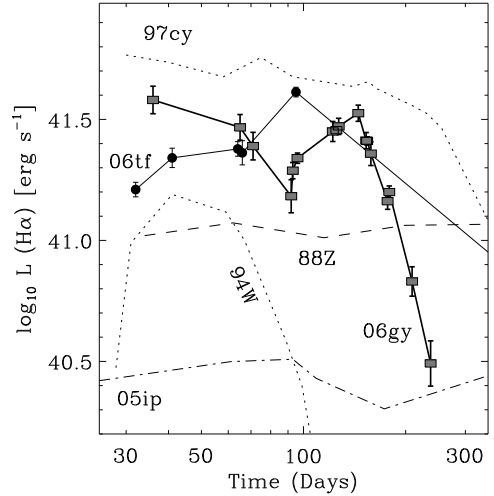}
\caption{Observed H$\alpha$ luminosity of SN~2006gy, corrected for
  Galactic and local extinction, compared to that of several other SNe~IIn
  (from Smith et al.\ 2008a, 2009a).}
\label{fig:haLUM}
\end{figure}

The measured H$\alpha$ equivalent width and line luminosity are shown
in Figures~\ref{fig:haEWgy}, \ref{fig:haEW}, and \ref{fig:haLUM}, and are
discussed below.  The measured H$\alpha$ equivalent width has several
potential sources of error.

First, we are comparing data obtained with different telescopes,
instruments, and (most importantly) slit widths.  This is not
a severe problem at early times when the SN is bright, but it can be
problematic at late phases.  Figure~\ref{fig:haEWgy} shows the
equivalent width plotted with different symbols corresponding to the
three spectrographs we used.  Spectra obtained at Lick Observatory
have the largest slit width, so these equivalent width measurements
are likely to have the largest systematic error because of the
difficulty in subtracting background light.  Overall, we found that
the LRIS and DEIMOS measurements were very consistent, whereas the
Lick measurements at comparable epochs showed larger variation.  For
this reason, we adjusted the equivalent withs of the Lick measurements
to correct for partial contamination by 10$^{-15}$ erg s$^{-1}$
cm$^{-2}$ \AA$^{-1}$ of galaxy continuum light in the slit, which
brought the late-time equivalent width into agreement with the LRIS
and DEIMOS values.  This is justified by the similar line luminosities
at those times, which are not altered by the continuum level.

Second, the H$\alpha$ profile is complex, with several adjacent
overlapping line features that change with time.  For example, at late
times, the red wing of H$\alpha$ descends into a broad absorption
feature at 6700--7000 \AA, which alters the underlying continuum
level.  The blue wing is plagued by broad P Cygni absorption and by
Fe~{\sc ii} and Si~{\sc ii} lines.  These complexities are exacerbated
by the different spectral resolutions of different instruments.  We
aimed to consistently measure the H$\alpha$ equivalent width and flux
over a wavelength range of 6400--6700 \AA, although the effects noted
above introduce uncertainty in the underlying continuum level at
5--10\%.  Equivalent width and flux measurements listed in Table~1
reflect the total H$\alpha$ line, including intermediate-width and
narrow components.  We made no correction for Fe~{\sc ii} and 
He~{\sc i} lines included in the sampled wavelength range.

\subsubsection{H$\alpha$ Equivalent Width}

The H$\alpha$ equivalent width changes dramatically as SN~2006gy
evolves, decreasing rapidly during the first 100 days, and then rising
steeply for the next $\sim$50 days, only to drop off again at late
times (Fig.~\ref{fig:haEWgy}).  The H$\alpha$ equivalent width
appears to be roughly anticorrelated with the SN luminosity, with 
the minimum equivalent width occurring near peak.  The solid line in
Figure~\ref{fig:haEWgy} demonstrates what the equivalent width would
be if the H$\alpha$ line luminosity were to remain constant with time.
This gives a fairly good match to the observed behavior during the
main light-curve peak, considering that the H$\alpha$ line is affected
by strong P Cygni absorption that diminishes the equivalent width by
varying amounts with time.  This implies that the production of
H$\alpha$ is not closely tied to the engine that drives the
continuum luminosity of SN 2006gy.

The H$\alpha$ equivalent width of SN 2006gy is compared to that of
several other SNe~IIn in Figure~\ref{fig:haEW}; we find that SN~2006gy
behaves unlike any of them.  All SNe~IIn show a consistent trend of
steadily increasing H$\alpha$ equivalent width with time, presumably
due to steadily decreasing optical depth in the post-shock shell.
Unlike the case in other SNe~IIn, the H$\alpha$ equivalent width of SN
2006gy plummets at first, rises sharply, and then drops off again. We
conjecture that during early phases, the photosphere is ahead of the
shock, but the amount by which the photosphere leads the shock
decreases with time. The difference compared to other SNe~IIn is
probably a consequence of SN~2006gy's much slower rise to peak
luminosity because of much higher optical depths --- in other words,
similar phases in other SNe~IIn may occur quickly at early times and
may be rarely observed.

The subsequent rise in equivalent width during intermediate phases
(days 90--150) seems consistent, at least for this brief interlude,
with the behavior of other SNe~IIn.  This phase starts just before the
critical time when $R_{\rm BB}$ turns over and the shocked shell
starts to become effectively optically thin.  This seems to imply that
during this phase, radiation from CSM interaction in SN~2006gy behaves
like in standard SNe~IIn because by this time the post-shock layer is
becoming sufficiently optically thin to cool directly.

However, in addition to behaving differently with time, the measured
H$\alpha$ equivalent width of SN 2006gy is also much lower than that
of any other SN~IIn in Figure~\ref{fig:haEW}.  At its time of maximum
luminosity $\sim$70 days after explosion, the H$\alpha$ equivalent
width of SN~2006gy is 20 times lower than in a standard SN~IIn like
SN~1998S, and almost a factor of 10 lower than in the very luminous
SN~IIn 2006tf.  This implies that direct cooling from ongoing CSM
interaction contributes only a fraction of the total luminosity. (Note
that the weakness or lack of He~{\sc i} suggests that the H abundance
is not particularly low.)

\subsubsection{H$\alpha$ Line Luminosity}

Examination of the intrinsic H$\alpha$ luminosity sheds additional
light on the problem (Fig.~\ref{fig:haLUM}).  In agreement with our
conjecture from the equivalent width, the H$\alpha$ line luminosity
stays roughly constant during the first 150 days: it does not change
by more than a factor of 2 while the SN continuum luminosity brightens
and fades by a factor of 10.  The relative minimum around day 100
coincides with a time when the broad P Cygni absorption was
particularly strong.

Figure~\ref{fig:haLUM} reveals that SN~2006gy has similar H$\alpha$
line luminosity compared to other luminous SNe~IIn.  The total
H$\alpha$ luminosity hovers around (2--3) $\times 10^{41}$ erg
s$^{-1}$, comparable to that of SN~1997cy and SN~2006tf.  This is
somewhat surprising, as SN~2006gy was a factor of 5--10 more luminous
in the continuum.  Together, then, the H$\alpha$ equivalent width and
luminosity imply that the engine that generates the H$\alpha$ line
luminosity (probably direct cooling from post-shock gas at late times)
may not be the dominant source of the peak continuum luminosity.  One
likely solution is that the continuum luminosity includes a
contribution from shock-deposited thermal energy that occurred earlier
and must diffuse out of the opaque shocked shell after a delay (Smith
\& McCray 2007).

At late times after day 150, both the equivalent width and total
H$\alpha$ luminosity drop.  The closest behavior to this is in
SN~1994W, which also showed a sharp drop, while SN~1997cy dropped off
at later times as well.  As in the case of SN~1994W (Chugai et al.\
2004), the most likely explanation is that the blast wave reached an
outer boundary of a circumstellar shell.  This suggests, therefore,
that the CSM had a sharp density drop at a radius of roughly
10$^{16}$~cm or $\sim$600~AU.  Again, however, we see that the
continuum luminosity did not exhibit such a sharp drop at that time,
suggesting that ongoing CSM interaction was only contributing a
portion of the total continuum luminosity.

\subsubsection{The H$\alpha$/H$\beta$ Line Ratio}

Finally, in Table 1 we also list the dereddened H$\alpha$/H$\beta$
line intensity ratio.  The first few epochs have reliable
measurements, but after the time of peak luminosity the ratio becomes
less certain: H$\beta$ becomes extremely faint and blended with
neighboring broad features, so the continuum level is unclear.  Also,
it develops deep self absorption that sometimes overwhelms the
emission component.  At late times, it is difficult to discern any
intermediate-width H$\beta$ component at all.  H$\beta$ emission is
dominated at late times by a narrow feature from the pre-shock CSM.
By contrast, the narrow pre-shock CSM component of H$\alpha$ typically
contributes less than 10\% of the total line flux.

At early epochs before and during maximum luminosity, the
H$\alpha$/H$\beta$ line ratio is roughly consistent with
recombination, supporting the notion that the line is formed largely
by photoionization of pre-shock circumstellar gas and subsequent line
broadening by electron scattering.  At later epochs the
H$\alpha$/H$\beta$ line ratio is extremely high, with all values well
above 30, as seen in late-time spectra of SNe~IIn with strong
interaction like SN~1988Z and SN~2006tf (Turatto et al.\ 1993; Smith
et al.\ 2008a).  Even taking into account the large uncertainties,
this clearly suggests collisional excitation in a shock, and therefore
implies that CSM interaction dominates the H$\alpha$ luminosity during
the decline from maximum.  A very similar growth in the
H$\alpha$/H$\beta$ ratio was seen in SN~2005gj (Prieto et al.\ 2007),
although the transition from low to high values occurred more quickly
there.

The onset of high H$\alpha$/H$\beta$ ratios ($>30$) after day 100
appears to coincide with the time when $R_{\rm BB}$ turns over and the
H$\alpha$ equivalent width begins to grow, presumably marking the
point when the swept-up shell starts to become optically thin.  {\it
  From this time onward, the optically thin H$\alpha$ luminosity
  should therefore be a reliable indicator of the strength of CSM
  interaction and a tracer of the mass-loss rate of the progenitor
  star at large radii.}  This is a key point, since the H$\alpha$
luminosity is lower than one expects from the continuum luminosity
even at those late phases.

\begin{figure}
\epsscale{0.99}
\plotone{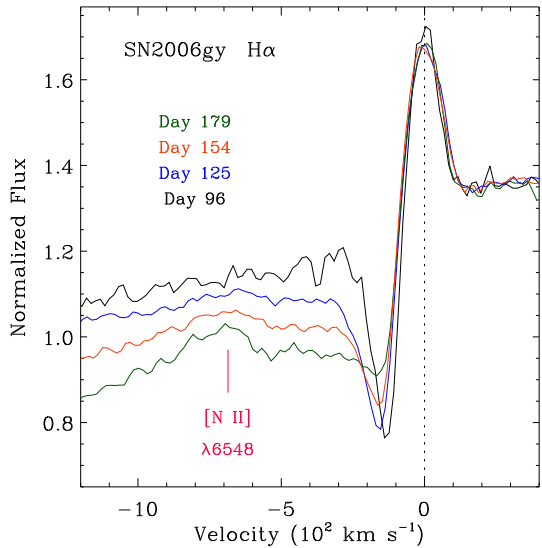}
\caption{The narrow component of the H$\alpha$ line profile in
  SN~2006gy on the four DEIMOS epochs observed at high resolution,
  normalized to the same continuum level.  The narrow blueshifted
  H$\alpha$ absorption gets weaker but broader with time.}
\label{fig:haNarrow}
\end{figure}

\begin{figure}
\epsscale{0.95}
\plotone{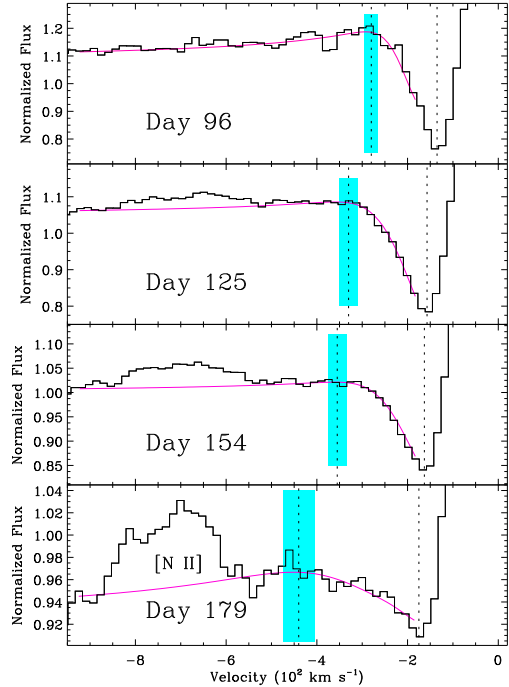}
\caption{Same data as in Figure~\ref{fig:haNarrow}, but zooming in on
  the narrow H$\alpha$ absorption on each DEIMOS epoch.  The purpose
  here is to show how we defined the blue-edge velocity of the narrow
  absorption by fitting a smooth curve (magenta) to the transition
  from the narrow absorption to the blue emission wing, excluding the
  narrow [N~{\sc ii}] $\lambda$6548 emission.  The point where this
  curve reaches a maximum is taken to be the blue-edge velocity, shown as
  the vertical dashed line, with uncertainties indicated by the blue box.
  The velocity of the minimum of the narrow absorption is also marked
  with a dashed vertical line.}
\label{fig:haAbs}
\end{figure}

\begin{figure}
\epsscale{0.99}
\plotone{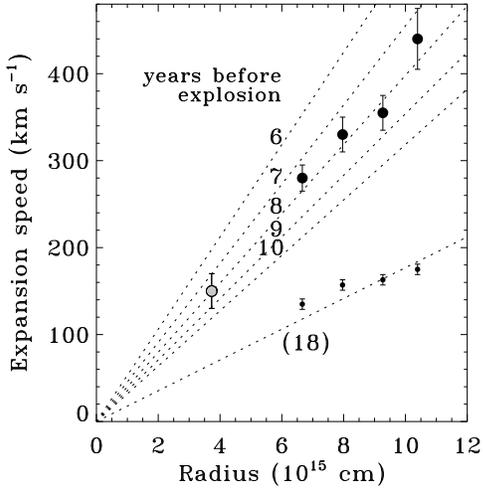}
\caption{Velocities measured from the narrow component of the
  H$\alpha$ line profile in SN~2006gy on the four DEIMOS epochs in
  Figure~\ref{fig:haNarrow}, plotted as a function of the presumed
  forward-shock radius at the same time (see text and
  Figure~\ref{fig:lc}).  The large solid points are the blue edge of
  the narrow absorption and the small points are the velocity for the
  absorption minimum.  The unfilled point corresponds to the +150 km
  s$^{-1}$ shift of the centroid of the Lorentzian fit to the day 36
  H$\alpha$ profile, representing the speed of the CSM outside the
  shock at that time (see \S 3.3.1 and Fig.~\ref{fig:haLor}).  The
  dashed lines represent the velocity one expects for simple Hubble
  laws ($V \propto R$) for CSM shell ejection times prior to the SN
  explosion in years.  (The $\sim$18 yr age that approximates the
  velocity profile of the absorption minimum is not really
  representative of the shell age, as the deepest absorption occurs
  away from the line of sight.)}
\label{fig:haVELnarrow}
\end{figure}

\subsection{The Narrow H$\alpha$ Line Profile}

In addition to the broad H$\alpha$ component discussed above,
SN~2006gy also has strong narrow H$\alpha$ emission from the
photoionized but unshocked CSM.  In Paper~I we presented a
high-resolution spectrum of H$\alpha$ on day 96, which revealed a
narrow blueshifted P Cygni absorption line as well.  This component
indicated outflow speeds of $\sim$260 km s$^{-1}$ in the pre-shock
CSM.  If this is attributed to the wind speed of the progenitor star,
it suggests an escape velocity comparable to that expected in blue
supergiants.  If it is associated with an explosive pre-SN ejection,
it is reminiscent of the speeds seen from LBV eruptions (see Paper~I).
It is much faster than speeds expected for a red supergiant (RSG), and
much slower than for a Wolf-Rayet star.

Figure~\ref{fig:haNarrow} shows the narrow H$\alpha$ absorption and
emission feature in SN~2006gy at several epochs as seen in our
high-resolution DEIMOS spectra.  Two qualitative points are evident
from this plot.

(1) The strength of the blueshifted absorption weakens relative to the
narrow emission component.  On day 96 the two are comparable, implying
that it may be a pure scattering P Cygni profile, but as time proceeds,
this narrow blueshifted absorption weakens.

(2) The velocities of the absorption minimum and the blue edge of the
narrow absorption feature shift to faster speeds at later times.  The
speeds of the absorption minimum are roughly $-$135, $-$157, $-$163, and
$-$175 ($\pm$5) km s$^{-1}$ on days 96, 125, 154, and 179, respectively.
The blue-edge velocities are less clear, but are roughly 280($\pm$15),
330($\pm$20), 355($\pm$20), and 440($\pm$35) km s$^{-1}$ on these same
days.  We define the blue-edge velocity as the point where the
intensity turns over and begins to drop down the slope of the broad
component, as shown in Figure~\ref{fig:haAbs}.  It is complicated by
the presence of [N~{\sc ii}] $\lambda$6548 and by noise or complex
structure in the data, so these values are approximate.  Nevertheless,
it is clear that the velocity changes with time in a systematic way.
Kawabata et al.\ (2009) obtained high-resolution spectra ($R \approx
3600$) on days 127 and 157, and their analysis found that the narrow
absorption developed a double component and widened to 690 km
s$^{-1}$.  This is not confirmed by our spectra at similar epochs with
higher resolution, although we note that we can reproduce a similar
increase in absorption velocities if we oversubtract neighboring
background H~{\sc ii} region emission.

The changing depth of the narrow absorption probably results from the
simple fact that as the blast-wave radius increases with time and
sweeps through the CSM, the total column density of the pre-shock gas
is decreasing.  This occurs because the immediate pre-shock density
should drop with increasing radius, and because increasing amounts of
CSM are already swept up behind the shock.  This is discussed in more
detail in \S 3.6, in context with the behavior of other narrow
absorption lines in the spectrum.

The increasing speed seen in the narrow blueshifted absorption
components (Fig.~\ref{fig:haAbs}) is more interesting, as it directly
samples the velocity profile of the pre-SN mass loss.  We can use
these observed velocities to reconstruct this pre-SN velocity profile
and thereby infer the age of the CSM.  If we assume that the narrow
absorption at each epoch traces cool pre-shock material just outside
the forward shock at that time, we can derive its approximate radius
away from the progenitor by using the radius given in the bottom panel
in Figure~\ref{fig:lc}.  (We also assume that the blast wave continues
to expand at $\sim$5200 km s$^{-1}$ after day 110 when the derived
value of $R_{\rm BB}$ decreases due to effective optical depth, as
explained in \S 3.2.)  Calculating the radius in this way, we plot
corresponding values of $R$ and $V$ in Figure~\ref{fig:haVELnarrow}.

The blue-edge velocities are probably the best indicator of the
outflow speed of pre-shock gas along the line of sight at each epoch,
whereas the absorption minimum probably traces velocities with some
projection from the line of sight closer to the limb of the shell.  We
see clearly that the velocity of pre-shock gas along the line of sight
increases with radius.  We also plot the CSM speed of 150 km s$^{-1}$
inferred from the shifted centroid of the Lorentzian fit to the day 36
H$\alpha$ profile, assuming that it is caused by electron scattering
(see \S 3.3.1). For comparison in Figure~\ref{fig:haVELnarrow}, we
also plot dashed lines corresponding to the velocity profile one
expects for simple Hubble laws ($V \propto R$) for CSM shell ejections
that occurred 6, 7, 8, 9, and 10 yr prior to the SN explosion.
Although the velocities do not perfectly match a linear increasing
speed with increasing radius, they are consistent with such a trend
within the uncertainties.  The age of the CSM shell appears to be
$7.8 \pm 1$ yr before explosion, if the expansion has been roughly
constant, implying that the massive pre-SN shell ejection experienced
by the progenitor of SN~2006gy occurred around October to December
1998.


There are several interesting implications.  A Hubble law in the CSM
velocity profile means that the observed speeds are {\it not} the
consequence of acceleration of the pre-shock gas by the radiation
force of the SN luminosity itself; that mechanism is expected to cause
a negative velocity gradient (e.g., Chugai et al.\ 2002), which is the
opposite of the observed trend.  Perhaps the pre-shock CSM speeds of
other SNe~IIn, such as 400, 700, and 1000 km s$^{-1}$ around
SNe~1998S, 1995G, and 1994W, respectively (Chugai et al.\ 2002, 2004;
Chugai \& Danziger 2003), may reflect their progenitors' ejection
speeds, rather than radiative acceleration by the SN.

If the CSM speed really is tied to the physics of the progenitor's
mass ejection, rather than having been radiatively accelerated
afterward by the SN, it implicates a progenitor object with an escape
speed significantly faster than a large RSG, as we discussed in
Paper~I.  Instead, the observed Hubble-like velocity profile points to
a relatively sudden episode of mass ejection with speeds of several
hundred km s$^{-1}$.  Regardless of the cause, this is similar to the
range of velocities and Hubble-like profiles observed in the ejecta of
massive LBVs such as $\eta$~Carinae (Smith 2006a), although the ejecta
shell around SN~2006gy was much younger and therefore much closer to
the star.  In Paper~I we noted speculative connections between
the precursor event of SN 2006gy and the 1843 eruption of $\eta$~Carinae
based on the H-rich composition, total energy of $\sim$10$^{50}$ erg,
and ejected mass of order $\sim$10 M$_{\odot}$ (Smith et al.\ 2003),
and the ejecta speed.  The Hubble-like velocity laws seen in both
objects reinforces this comparison.  It is also a feature that is
explicitly predicted in the model of Woosley et al.\ (2007), where the
pre-SN mass loss is caused by a $\sim$10$^{50}$ erg pulsation or
explosion triggered by the pair instability. In fact, the observed
delay time of 7--9 yr between the precursor and SN~2006gy is in good
agreement with the 6.8 yr delay predicted in Woosley et al.'s model.
The Hubble-like profile is not consistent with a dense but steady
mass-loss wind, indicating that the progenitor star was decidedly
unstable leading up to core collapse.  For SN~2006gy, the precursor
event ejected $\sim$19 M$_{\odot}$ at speeds of 200--500 km s$^{-1}$,
with a corresponding kinetic energy of $\ga$10$^{49}$ erg.

The fact that the systematically increasing CSM velocity is seen in
absorption means that it corresponds to material along our line of
sight.  The range of velocities is therefore not likely to be caused
by asymmetry in the CSM, although by no means is this an argument that
the CSM is necessarily spherical.  We hope to investigate possible
effects of asymmetry in a future paper discussing our
spectropolarimetry of SN 2006gy.  Asymmetry in the CSM is appealing as a
potential explanation for the very smooth and slowly evolving light
curve (van Marle, Smith, \& Owocki 2009b).

\begin{figure*}
\epsscale{0.9}
\plotone{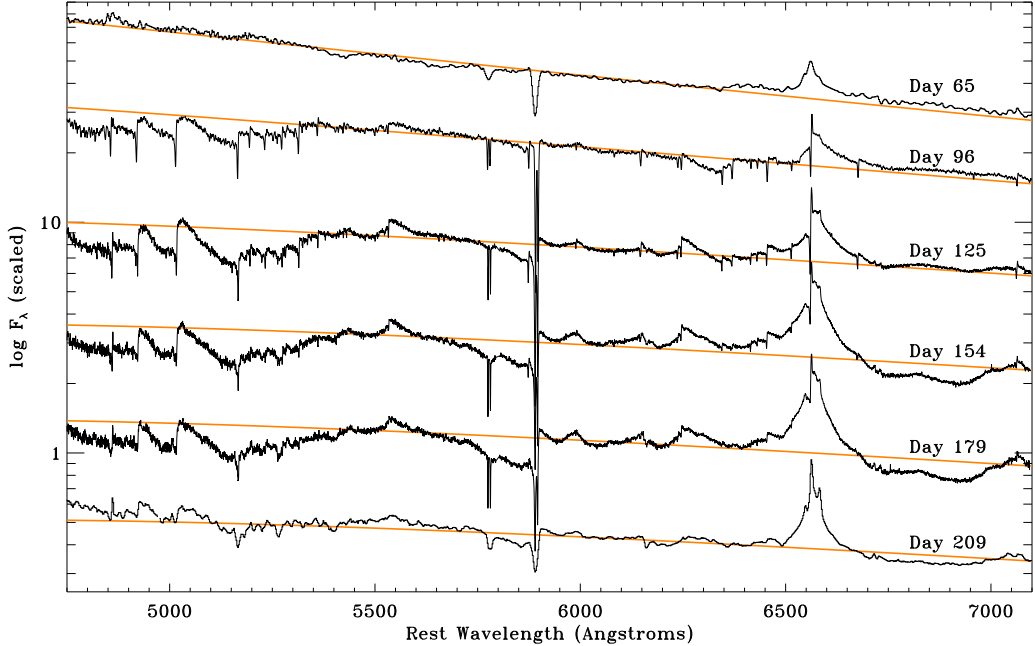}
\caption{Same as Figure~\ref{fig:dered}, but displaying a smaller
  wavelength range and emphasizing our high-resolution DEIMOS spectra
  on days 96, 125, 154, and 179, as well as one epoch before and after
  with lower resolution. These have been dereddened by correcting for
  both Galactic and host-galaxy extinction as before, with the same
  underlying blackbodies for comparison (gray; orange in the online
  edition).  A key point is that the many narrow absorption
  components weaken with time after day 96.}
\label{fig:zoom}
\end{figure*}

\subsection{Peculiar Narrow and Broad Absorption Features}

Figure~\ref{fig:zoom} gives a detailed view of our high-resolution
spectra of SN~2006gy at visual wavelengths from the blue to H$\alpha$,
which emphasizes features that make the spectrum of SN~2006gy so
distinct.  The most remarkable characteristic of the spectrum
is the presence of line profiles with both broad and narrow emission
and absorption components, leading in some cases to profiles with
extremely sharp edges that separate redshifted emission from
blueshifted absorption.  As we noted earlier in Figure~\ref{fig:comp},
the narrow absorption features in SN~2006gy find their closest
comparison with the spectra of SN~1994W and SN~1995G (Sollerman et
al.\ 1998; Chugai \& Danziger 2003; Chugai et al.\ 2004).  However,
the dual nature of the SN~2006gy profiles --- with clearly separate
broad and narrow absorption components --- was not seen in those two
SNe.  These lines are not as prominent in any other known SN~IIn.

A key point to recognize in Figure~\ref{fig:zoom} is that after first
appearing in the day 96 spectrum (and the days 92 and 93 spectra in
Figs.~\ref{fig:dered}), the narrowest absorption features become
systematically weaker with time.  These narrow absorption features at
velocities of roughly $-$200 km s$^{-1}$ are formed in the pre-shock
CSM.  Their progressive weakening is {\it not} an effect
of increasing or decreasing ionization with time in the pre-shock CSM,
since the same trend is seen in narrow absorption of both He~{\sc i}
and Fe~{\sc ii} lines with widely different ionization
state.\footnote{The four narrow Na~{\sc i} absorption lines are in a
separate category, as they are formed largely in the Milky Way 
interstellar medium (ISM) and in the ISM of NGC~1260, rather than in 
the immediate CSM of the progenitor of SN~2006gy.}

This effect signals that the column density of pre-shock absorbing
material is probably decreasing with time as the forward shock
approaches the outer boundary of a dense shell.  The weakening of the
narrow absorption features occurs from days 90 to 200; many of the
narrow features are barely discernible on day 179.  Recall (\S 3.4.2)
that the total line luminosity of H$\alpha$ begins to drop around day
150 and plummets severely by day 209 (Figure~\ref{fig:haLUM}).  The
H$\alpha$ emission line is produced primarily in the post-shock gas,
which is becoming optically thin at these late phases, so it directly
traces the strength of the CSM interaction and hence, the pre-shock
CSM density.  The fact that both this post-shock H$\alpha$ emission
and the pre-shock narrow absorption features weaken at the same time
confirms the notion that by day 200, the blast wave of SN~2006gy is
overtaking the outer boundary of a shell that was probably ejected
$\sim$8 yr before core collapse.

\begin{figure}
\epsscale{0.99}
\plotone{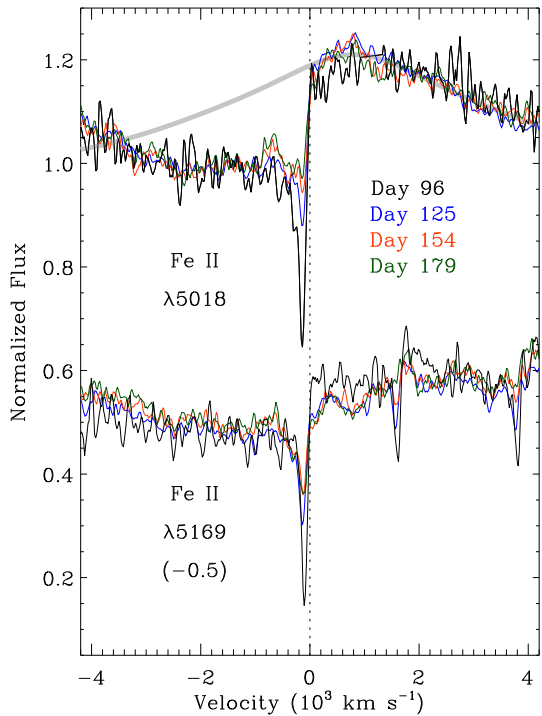}
\caption{The observed profiles of Fe~{\sc ii} $\lambda$5018 in
  high-resolution DEIMOS spectra.  As with H$\alpha$, the narrow
  absorption becomes relatively weaker with time.  The thick grey
  curve is a 2-component composite Gaussian with the same FWHM
  velocities as for H$\alpha$, but with different strengths and with
  its center offset by $+$600 km s$^{-1}$.  We also show tracings of
  Fe~{\sc ii} $\lambda$5169 for comparison.}
\label{fig:fe}
\end{figure}

{\bf Fe~II $\lambda$5018 and $\lambda$5169:} Figure~\ref{fig:fe} shows
the profiles of Fe~{\sc ii} $\lambda$5018 observed at high resolution.
This is a strong line that is representative of many other features in
the spectrum.  The red wing of its smooth broad emission component
indicates velocities similar to those of H$\alpha$, suggesting that it is
formed in a similar region of the swept-up CDS.  It is not present in
the early-time spectrum before and during peak luminosity, when we
expect that the photosphere is ahead of the forward shock, but the
broad Fe~{\sc ii} $\lambda$5018 emission is present at all times after
the post-shock gas becomes optically thin. Figure~\ref{fig:fe}
demonstrates the extremely sharp transition from redshifted emission
to blueshifted absorption that is seen in many lines in the spectrum
of SN~2006gy.  The gray curve in Figure~\ref{fig:fe} is a symmetric
Gaussian matching the red emission wing, showing that Fe~{\sc ii}
$\lambda$5018 suffers from significant broad blueshifted
self-absorption out to $-$4000 km s$^{-1}$, as seen in H$\alpha$.
This broad absorption remains constant with time.  Fe~{\sc ii}
$\lambda$5169 has a similar broad absorption feature, but lacks the
broad emission.

Relative to the broad-emission strength, the narrow absorption from
the pre-shock CSM weakens with time as noted above for many other
lines.  Its equivalent width decreases by more than a factor of 2 from
day 96 to 179.  The velocity of the pre-shock gas causing the narrow
Fe~{\sc ii} $\lambda$5018 absorption does not show as clear of an
increase with time as had been seen in H$\alpha$, but the day 96
spectrum is noisy for these Fe~{\sc ii} lines so it is difficult to
evaluate changes in the narrow absorption profile.  Fe~{\sc ii}
$\lambda$5169 has a narrow absorption feature that is very similar to
that of Fe~{\sc ii} $\lambda$5018 (Fig.~\ref{fig:fe}), so we infer that 
the Fe~{\sc ii} $\lambda$5018 line is not severely affected by possible
He~{\sc i} $\lambda$5015.  The narrow Fe~{\sc ii} absorption indicates
speeds in the CSM of 200--600 km s$^{-1}$, consistent with H$\alpha$.
The more subtle velocity evolution may suggest that the narrow CSM
absorption of Fe~{\sc ii} occurs over the full range of radii in the
CSM shell, whereas H$\alpha$ absorption occurs preferentially over a
smaller range of radii near the blast wave.  This could be a result of
the more demanding conditions for nebular Balmer absorption as opposed
to those needed for Fe~{\sc ii}.  To fully understand the differences
between Fe~{\sc ii} and H$\alpha$ absorption and emission,
time-dependent radiative transfer models are needed.

\begin{figure}
\epsscale{0.99}
\plotone{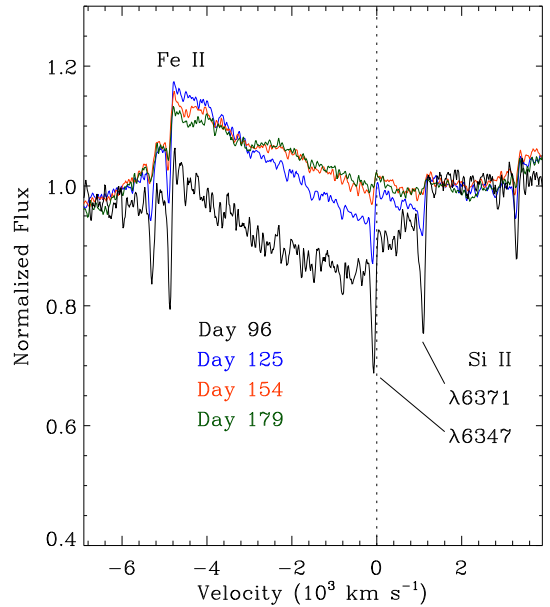}
\caption{DEIMOS spectra at wavelengths surrounding the red Si~{\sc ii}
  doublet, plotted on the velocity scale for Si~{\sc ii}
  $\lambda$6347.  In addition to narrow Si~{\sc ii} absorption,
  possible broad absorption out to a few thousand km s$^{-1}$ is seen
  on day 96, but its interpretation is complicated because of blending
  with a Fe~{\sc ii} emission feature that strengthens with time.}
\label{fig:si}
\end{figure}

{\bf Si~II $\lambda\lambda$6347, 6371:} Figure~\ref{fig:si} shows the
profiles of the Si~II $\lambda\lambda$6347, 6371 doublet observed at
high resolution.  These lines are of particular interest since they
are a defining characteristic of SNe~Ia, and the presence of this
broad absorption feature in spectra of SN~2006gy could suggest a
possible connection to objects like SN~2002ic and 2005gj, although as
noted elsewhere (Ofek et al.\ 2007; Paper~I), these lines are also
seen in some core-collapse SNe.

The Si~{\sc ii} $\lambda\lambda$6347, 6371 doublet exhibits narrow
absorption that weakens with time, in agreement with several other
absorption lines formed in the pre-shock CSM.  The doublet shows no
narrow or broad emission components.

More interesting, perhaps, is the broad blueshifted Si~{\sc ii}
absorption in Figure~\ref{fig:si}.  The shape of the blue wing is
compromised by a blend with Fe~{\sc ii}, but it is clear that
significant broad blueshifted absorption of Si~{\sc ii} is present on
day 96, out to velocities of roughly $-$4000 km s$^{-1}$.  This
absorption weakens with time, perhaps due to ionization effects, while
it is weak or absent in early-time spectra.  The speeds traced by this
blueshifted Si~{\sc ii} are far too slow to arise in a normal SN~Ia
atmosphere, where speeds well above 10$^4$ km s$^{-1}$ are seen, but
velocities as fast as 4000 km s$^{-1}$ are entirely consistent with
speeds we have inferred for the expanding post-shock shell.  Similar
broad Si~{\sc ii} absorption was seen in SN~1998S (Fassia et al.\
2001).  We conclude that the broad Si~{\sc ii} absorption in SN~2006gy
is not formed in the photosphere of the underlying SN, but is instead
in the post-shock shell, and should therefore not be taken as
providing any connection to a putative SN~Ia.  As noted elsewhere, the
post-shock shell is opaque until day 110, so features associated with
the underlying SN photosphere should not be seen before then anyway.

\begin{figure}
\epsscale{0.99}
\plotone{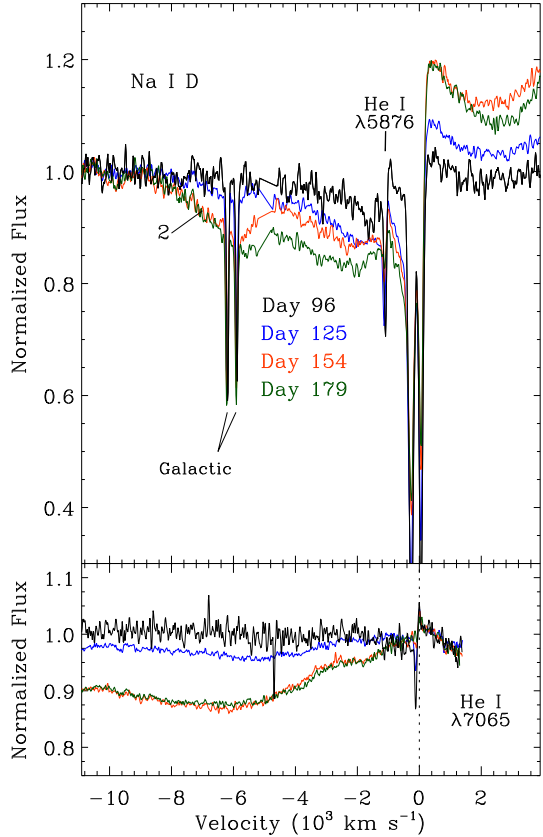}
\caption{{\it Top}: The observed line profiles near Na~{\sc i}~D in
  high-resolution DEIMOS spectra, including broad emission and
  absorption of Na~{\sc i} (or He~{\sc i}), Galactic and host-galaxy narrow
  absorption from Na~{\sc i}, plus weak narrow absorption and emission
  from He~{\sc i} $\lambda$5876. {\it Bottom}: Tracings of He~{\sc i}
  $\lambda$7065 in the same data for comparison.}
\label{fig:sodium}
\end{figure}

{\bf Na~I D and He~I $\lambda$5876:} The most striking features
associated with the Na~{\sc i} line in Figure~\ref{fig:sodium} are the
four very narrow absorption features that correspond to the two pairs
of Na~{\sc i} absorption lines formed in the Milky Way ISM and in the
ISM of NGC~1260.  These features do not change much with time when
observed at consistent resolution, and support the large values of
Milky Way and host-galaxy reddening that we inferred in Paper~I.
Figure~\ref{fig:sodium} also reveals narrow absorption and emission
from He~{\sc i} $\lambda$5876 (and $\lambda$7065), consistent with
other narrow He~{\sc i} features in the spectrum.

The identification of the broad emission in the top panel of
Figure~\ref{fig:sodium} is unclear. Due to the proximity in wavelength
of Na~{\sc i} D and He~{\sc i} $\lambda$5876, it could be either or
both at different times.  It follows the same trend of increasing
equivalent width of H$\alpha$ emission over the same time period.
Since Na~{\sc i} is expected to be strong under similar conditions in
SN~II spectra (e.g., Dessart \& Hillier 2008) and because similar
increases in strength are not seen in emission components of other
He~{\sc i} lines (like He~{\sc i} $\lambda$7065; bottom of
Fig.~\ref{fig:sodium}), we surmise that the broad {\it emission}
component is Na~{\sc i}.

The nature of the broad {\it absorption} in Figure~\ref{fig:sodium} is
more perplexing.  This absorption strengthens with time, whereas
Fe~{\sc ii} $\lambda$5018 stays roughly constant and Si~{\sc ii}
decreases in strength with time.  Unlike any broad absorption features
discussed so far, it reaches to velocities as fast as $-$8000 km
s$^{-1}$ on days 154 and 179.  The broad absorption blueward of
He~{\sc i} $\lambda$7065 behaves very differently in velocity and time
(Fig.~\ref{fig:sodium}, bottom), so we conjecture that {\it both}
features cannot be He~{\sc i} absorption; perhaps neither are.

The higher-speed absorption near Na~{\sc i} may be due to a second
absorbing species that we refer to here as component 2 (see
Figure~\ref{fig:sodium}).  The relative contribution of component 2
changes with time; it is initially absent and then becomes prominent
on days 154 and 179, forming an additional dip at velocities of
$-$5000 to $-$8000 km s$^{-1}$ that was absent before day 154.  It
seems likely that this additional absorption is due to a superposed
line of another species, since (a) the strength of component 2 changes
differently from the rest of the broad absorption feature, (b) it is
not seen in any other broad absorption line, and (c) fast SN ejecta
with speeds of $\sim$8000 km s$^{-1}$ should have already caught up
with the reverse shock by day 80 (see \S 4.3).  It may, however, be
related to other broad absorption features at red wavelengths that
strengthen at the same time, like the broad absorption blueward of
He~{\sc i} $\lambda$7065.

\begin{figure}
\epsscale{0.99}
\plotone{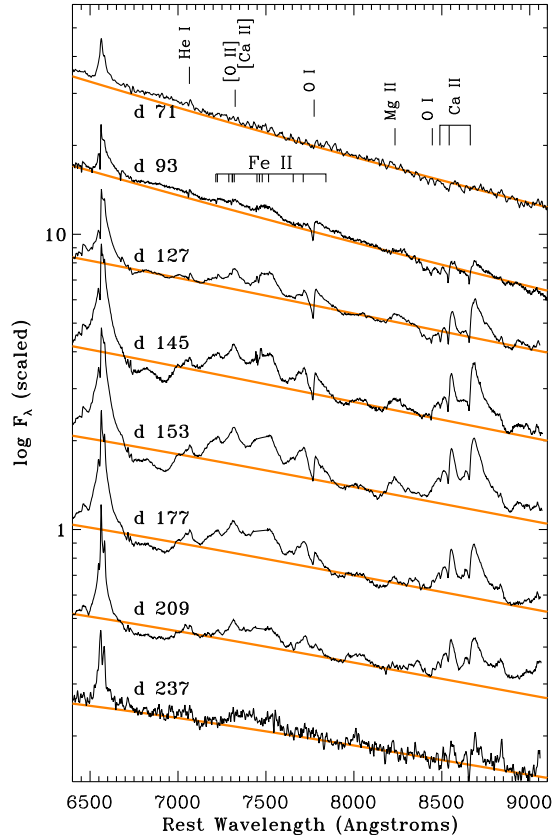}
\caption{LRIS spectra at red wavelengths, plus one Lick/Kast spectrum
  at peak luminosity for comparison.  Several line identifications are
  marked (see Dessart et al.\ 2009).  Representative blackbodies are
  shown (gray; orange in the online edition) as in
  Figure~\ref{fig:dered}.}
\label{fig:red}
\end{figure}

\begin{figure}
\epsscale{0.99}
\plotone{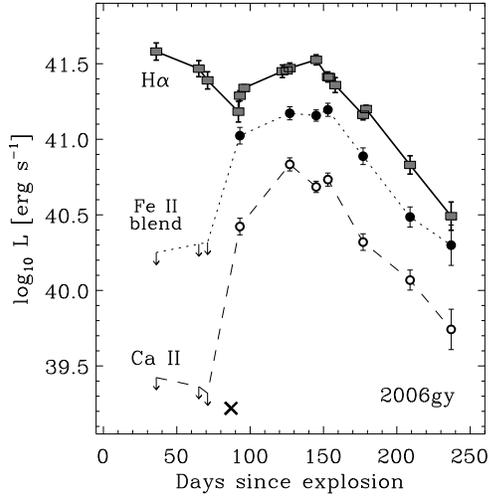}
\caption{The luminosity of the H$\alpha$ line from
  Figure~\ref{fig:haLUM} is compared with the integrated luminosities of
  the red Fe~{\sc ii} blend over roughly 7130--7620 \AA\ (filled
  circles) and that of Ca~{\sc ii} $\lambda$8662 (unfilled circles).
  The first three points for both Fe~{\sc ii} and Ca~{\sc ii} are
  generous upper limits; these features are not seen in the spectrum
  at early times.  The ``X'' marks the soft X-ray luminosity of SN 2006gy
  detected by {\it Chandra} that we reported in Paper~I.}
\label{fig:redLUM}
\end{figure}

\subsection{LRIS Spectra at Red Wavelengths}

We obtained six epochs of high signal-to-noise ratio (S/N) spectra
with LRIS on Keck (plus a last epoch on day 237 with lower S/N; see
Table 1).  These are particularly useful for investigating the
evolution of the red-wavelength spectrum of SN~2006gy
(Fig.~\ref{fig:red}), not sampled by our high-resolution DEIMOS data.
The day 71 spectrum is included in Figure~\ref{fig:red},
representative of our three early epochs of Lick/Kast spectra on days
36, 65, and 71, which exhibit essentially featureless blackbody
continua at red wavelengths with perhaps some weak absorption in the
Ca~{\sc ii} near-IR triplet and weak emission of He~{\sc i}
$\lambda$7065.

As shown by our sequence of LRIS spectra in Figure~\ref{fig:red}, the
post-maximum red-wavelength spectrum evolves with time.  Prominent
emission features are He~{\sc i} $\lambda$7065, a blend of several
Fe~{\sc ii} emission lines, O~{\sc i} $\lambda$7774, and the Ca~{\sc
  ii} near-IR triplet.  The last of these is particularly interesting;
it becomes nearly as strong as H$\alpha$, suggesting that it may be a
significant source of post-shock cooling in SNe~IIn.  A possible blend
of [O~{\sc ii}] and [Ca~{\sc ii}] near 7325 \AA \ also appears to be
present, although we cannot be confident of the identification of
these forbidden lines because they overlap with several Fe~{\sc ii}
lines.  The spectra in Figure~\ref{fig:red} are roughly consistent
with previous reports, but have better temporal sampling and
wavelength coverage.  Agnoletto et al.\ (2009) noted that the Ca~{\sc
  ii} triplet was present in their spectrum on day 174; our spectra
show that this was after the time when this line was strongest and was
already in a steady decline.  Spectra on days 127 and 157 presented by
Kawabata et al.\ (2009) showed He~{\sc i} $\lambda$7065 and a few of
the Fe~{\sc ii} emission lines out to $\sim$7400 \AA, although in our
spectra near those dates, these features appear stronger relative to
the continuum.  Lines in our red spectra during days 100--240 do not
correspond well with those in later spectra on day 394 identified by
Kawabata et al.\ (2009).

Our spectra reveal that the flux of these features changes notably
with time.  In Figure~\ref{fig:redLUM} we compare the temporal
evolution of the integrated luminosity of the Fe~{\sc ii} blend and
that of Ca~{\sc ii} $\lambda$8662 with the H$\alpha$ luminosity (from
Figure~\ref{fig:haLUM}).  To represent the behavior of the Fe~{\sc ii}
blend while avoiding the strong O~{\sc i} $\lambda$7774 absorption, we
took the flux above the continuum as it appears in
Figure~\ref{fig:red} measured over $\sim$7130--7620 \AA.  For Ca~{\sc
  ii} $\lambda$8662, we measured the line strength integrating over
roughly $-$2,000 to $+$4,000 km s$^{-1}$.  These emission features are
not seen in early-time spectra, so for those epochs we give
conservative upper limits.

Figure~\ref{fig:redLUM} shows a very interesting trend. The Fe~{\sc
  ii} and Ca~{\sc ii} emission track the behavior of the H$\alpha$
luminosity almost exactly after day 90, but at earlier times they are
relatively much fainter than H$\alpha$.  The likely implication is
that the Fe~{\sc ii} lines and the Ca~{\sc ii} triplet trace emission
directly from the post-shock cooling shell that is the product of CSM
interaction, which is invisible at early phases because it is behind
the external photosphere.  The same is true for the intermediate-width
H$\alpha$ in post-maximum phases.  At early times, however, when the
post-shock H$\alpha$ is hidden behind the photosphere, the emergent
H$\alpha$ line is dominated by a completely different emission
mechanism.  As we proposed above, the intermediate-width H$\alpha$
emission at early times probably arises from photoionized pre-shock
circumstellar gas and subsequent scattering by thermal electrons. We
find, therefore, that red Fe~{\sc ii} lines and the Ca~{\sc ii}
triplet are more robust tracers of the direct CSM interaction
luminosity than H$\alpha$, since intermediate width H$\alpha$ can be
produced by a different mechanism.  The Ca~{\sc ii} lines have no
narrow emission at any epoch.

\begin{figure}
\epsscale{0.99}
\plotone{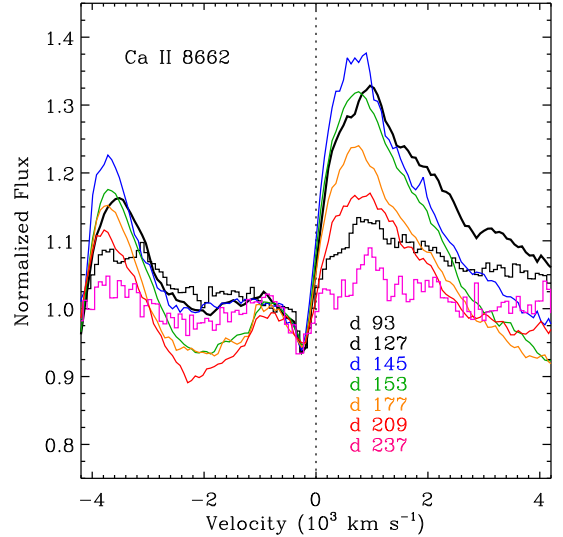}
\caption{LRIS spectra of Ca~{\sc ii} $\lambda$8662 at several epochs.
  These have been arbitrarily normalized at the minimum of the narrow
  absorption.  The first two phases we plot, shown in black, are while
  the line luminosity is increasing with time on days 93 (thin black
  histogram) and 127 (thicker black line).  After reaching its peak
  luminosity on day 145 (blue), the Ca~{\sc ii} $\lambda$8662 line
  weakens systematically with time (blue, green, orange, red, and
  magenta on days 145, 153, 179, 209, and 237 respectively).}
\label{fig:ca2}
\end{figure}

The red Fe~{\sc ii} lines are too blended to derive individual
line-profile shapes, but the profile of Ca~{\sc ii} $\lambda$8662 is
displayed in Figure~\ref{fig:ca2} as seen in all our epochs of LRIS
spectra.  Like H$\alpha$, Ca~{\sc ii} $\lambda$8662 has a bright
intermediate-width emission component with a half-width of roughly
2,000 km s$^{-1}$ and broader line wings.  Although the strength of
the line changes, its width does not change much.  Ca~{\sc ii}
$\lambda$8662 also has a sharp narrow blueshifted absorption component
at roughly $-$200 km s$^{-1}$, and is missing the blue side of the
line, suggesting rather strong broad blueshifted absorption.  Based on
the similarity of their red emission wings, H$\alpha$ and Ca~{\sc ii}
probably trace similar regions in the dense shell.

The presence of strong intermediate-width lines of Ca~{\sc ii} from
the post-shock gas is of potential interest with regard to SN~2008S
and the similar transient in NGC~300, both of which showed
particularly strong emission of H$\alpha$, [Ca~{\sc ii}], and the
Ca~{\sc ii} near-IR triplet (Smith et al.\ 2009b; Bond et al.\ 2009;
Berger et al.\ 2009).  SN~2006gy also may show the [Ca~{\sc ii}]
lines, although we cannot reliably separate them from [O~{\sc ii}] and
Fe~{\sc ii} lines.


Lastly, we note the presence of broad absorption features at
6700--7000 \AA, where the red wing of H$\alpha$ meets the continuum.
Signs of this absorption first appear on day 127, and the absorption
strengthens with time on days 145 and 153. It subsides at subsequent
epochs and disappears by day 237.  The change in strength of the broad
absorption features follows, in rough relative terms, the strength of
the emission in the red Fe~{\sc ii} and Ca~{\sc ii} lines noted above.
Prominent absorption at these wavelengths appears to be a common
feature in late-time spectra of luminous SNe~IIn such as SNe~1997cy,
1999E, 2002ic, and 2005gj.  For example, the day 98 spectrum of
SN~2005gj is shown in Figure~\ref{fig:comp}; it exhibits a single very
broad and prominent absorption feature at these wavelengths.  In
SN~2006gy this feature is either composed of two separate species with
absorption minima at $\sim$6750 and 6930 \AA, or alternatively, the
two parts are separated by a superposed emission feature at 6820
\AA.\footnote{Note that SN~2005gj exhibits an even stronger and
  broader absorption feature at $\sim$8000 \AA\ that is not seen in a
  comparable way in SN~2006gy at any epoch.  It is therefore possible
  that the 6700--7000 \AA\ broad absorption features in SNe~2005gj and
  2006gy are unrelated.}

The identification of these broad absorption lines is uncertain.
Broad lines near the wavelengths of these features are seen in SN
photospheres of Types~Ia, Ib, and Ic at late times, although the
different expansion speeds make a direct comparison with SN~2006gy
difficult.  This feature was part of the rationale for associating
SNe~2002ic and 2005gj with SNe~Ia, although Benetti et al.\ (2006)
have questioned this association and suggested SNe~Ic as the culprit
instead.  SN~2006gy informs this debate for the following reason: {\it
  In the case of SN~2006gy, these absorptions cannot be due to
  features in an underlying SN photosphere, because they are too
  strong.}

The fact that features are seen in absorption at such late phases
requires the presence of some substantial source of background
continuum light.  At its strongest, the broad absorption reaches a
depth of roughly 12--15\% of the continuum level
(Fig.~\ref{fig:redLUM}), requiring that the underlying source that is
being absorbed has a luminosity of (1--2) $\times 10^9$ L$_{\odot}$ at
the late phases of 120--170 days when this absorption is seen.  No
standard SN photosphere is that luminous at such late times, so these
lines cannot arise in a simple scenario of normal SN ejecta being seen
through the thinning CSM interaction region.  The luminosity may come
from the reverse shock or radioactive decay, as we discuss further in
\S 4.1.  More complicated effects may play a role as well, such as
``top-lighting'' (e.g., Branch et al.\ 2000), caused by irradiation of
the SN ejecta by photons emitted from the CSM interaction region.
Since SNe with strong CSM interaction like SN~2006gy can mimic
features seen in other SNe, this casts doubt on the association of
such features with an underlying SN~Ia photosphere in other luminous
SNe~IIn.  This would seem to support claims by Benetti et al.\ (2006)
that this class of objects could be SNe other than Type Ia.

Regardless of the identification, from the broad line profiles and the
fact that they do not appear until late times, it is likely that these
lines arise from SN ejecta that have just crossed or are just about to
cross the reverse shock.  As we note below, the speeds of unshocked SN
ejecta and the expansion speed of the CDS in the post-shock gas are
comparable at these late times.

\section{DISCUSSION}

\subsection{SN 2006gy Spectral Evolution Overview}

Altogether, the extraordinary spectral properties of SN~2006gy can be
understood in the context of CSM interaction where the radiating
layers evolved through three key phases in the first year, as follows.

(1) {\it Early times (days 0--90)}: At these phases before and up to
the time of maximum luminosity, the photosphere is outside the forward
shock and the spectrum is generated in the pre-shock CSM. Radiation
diffusing from the shocked shell propagates forward and pre-ionizes
the opaque CSM, leading to a smooth continuum plus Balmer lines
generated in CSM layers immediately outside that.  Consequently, the
post-shock CDS and the SN ejecta cannot be seen yet.  The
intermediate-width H$\alpha$ line profile arises because narrow lines
from the photoionized pre-shock CSM are broadened by multiple
scattering with thermal electrons in the opaque gas (e.g., Chugai
2001).  This hypothesis is consistent with the observed
H$\alpha$/H$\beta$ intensity ratio indicative of recombination at
early times, and the Lorentzian profiles of H$\alpha$ line wings.  At
this phase, therefore, the prime luminosity source is diffusion of
shock-deposited energy (Smith \& McCray 2007) in a manner similar to
that described by Falk \& Arnett (1977).

(2) {\it Decline from Maximum Luminosity (days 90--150)}: During these
phases, the photosphere recedes (in mass) past the forward shock and
into the CDS.  As the photosphere passes this sharp boundary, the
consequent transition in spectral morphology from phase 1 to phase 2
may be sudden.  The effective value of $R_{\rm BB}$ begins to fall,
not because the radius is actually decreasing, but because the CDS is
fragmented and is becoming progressively more transparent (Smith et
al.\ 2008a).  As the CDS is exposed, we begin to see strong emission
from cooling lines in the post-shock regions, such as multiple Fe~{\sc
  ii} and Ca~{\sc ii} lines, which turn on suddenly after day 90, in
addition to broad H$\alpha$.  Broad blueshifted absorption features
(see \S 4.2 below) appear during this time as the photosphere is now
inside the CDS, and outer layers of the CDS can be seen in absorption.
The broad emission-line wings and the blueshifted absorption show that
the expansion velocity of the CDS does not slow during this time
period, indicating self-similar expansion.  The broad line-wing shape
is probably a composite of the true kinetic line width (about 4000 km
s$^{-1}$ as indicated by the blueshifted absorption) plus a
contribution from electron scattering within the clumpy CDS itself
(not in dense CSM as in stage 1; see Dessart et al.\ 2009).

Much of the continuum luminosity during this time is still supplied by
fading diffusion from the gradually thinning shocked shell, possibly
reheated from within by reverse-shock luminosity, but there is also a
significant contribution now from ongoing CSM interaction luminosity.
The continuum effective temperature remains roughly constant.  The
speed of the pre-shock CSM appears to be increasing during this time,
suggesting a Hubble-like flow in the episodic pre-SN mass loss that
occurred $\sim$8 yr before explosion.  In addition, there is
suggestive evidence that during this phase we also see some broad
absorption features from the underlying SN ejecta that are about to
cross or have just crossed the reverse shock.  If true, the
interesting implication is that this absorption requires a fairly
substantial {\it background} continuum luminosity source of
$\sim10^{9}$ L$_{\odot}$.

(3) {\it Drop in Late-Time CSM Interaction (days 150--240)}: Beginning
around day 150, diagnostics of the post-shock cooling luminosity, such
as H$\alpha$, Fe~{\sc ii}, and Ca~{\sc ii} lines (see
Figure~\ref{fig:redLUM}), show a precipitous and continuing decline.
At these late phases the emitting layers must be optically thin and
getting thinner with time, so the weakness of these lines cannot be
attributed to high optical depth as before.  Instead, the decline
probably indicates that the CSM interaction luminosity itself is
dropping as the forward shock reaches the outer boundary of the dense
pre-SN shell ejected 8 yr earlier.  A key point is that during the
transition from stage 2 to stage 3, the narrow absorption components
formed in the pre-shock CSM were observed to weaken and disappear in
some cases, signaling lower pre-shock densities upstream, and
heralding the inevitable approach of the outer boundary of the dense
pre-shock CSM.  This confirms that the reason the H$\alpha$, Fe~{\sc
  ii}, and Ca~{\sc ii} line luminosities dropped at late times was
because the SN ran out of CSM to sweep up, not because the SN ran out
of energy or recombined.

The drop in CSM interaction at days 150--240 explains why the
late-time spectrum observed 1 yr later showed no sign of ongoing CSM
interaction.  H$\alpha$ was observed to be about 400 times weaker than
in SN~1988Z or SN~2006tf at a comparable epoch (Smith et al.\ 2008b).

Returning to the spectral behavior during stage 3 at days 150--240,
then, it remains quite perplexing why the continuum luminosity did not
drop by an amount comparable to the drop in H$\alpha$, Fe~{\sc ii},
and Ca~{\sc ii} as it did in SN~1994W (Chugai et al.\ 2004).  {\it It
  should have dropped if it were generated primarily by CSM
  interaction.}  Instead, the red-continuum luminosity appears to have
slowed its decline during this stage and even seemed to level off at
late times.  The SN was still quite bright at these late times --- it
was still as luminous as the peak of a normal Type Ia --- so this
extra late-time luminosity source after day 200 is nontrivial.

Given the difficulty imposed by the proximity of the host-galaxy
nucleus, it is conceivable that some of the blue-wavelength light
after day 200 is contaminated by host-galaxy light, but red
wavelengths show a clear detection of the SN.  In a companion paper
(Miller et al.\ 2009b), we confirm that the lingering emission
detected 1--2 yr later is indeed due to a light echo as proposed by
Smith et al.\ (2008b).  However, a reflected-light echo is not a
satisfying explanation for the extra luminosity around days 200--240,
because (1) the continuum shape observed then was redder than that at
the time of peak luminosity; we would expect reflected light to be the
same as or bluer than the original spectrum, as it is at later times
(Miller et al.\ 2009b), and (2) a distant light echo cannot give rise
to the broad absorption features seen in the spectrum (\S 3.7).

One possibility for the excess emission at days 200--240 is a
re-energized reverse shock, powered by increased density of SN ejecta
crossing the reverse shock and re-heating the CDS from the inside.
Even if the CSM density drops so that the forward shock no longer
encounters a significant obstacle, the massive CDS that has already
formed and is now coasting at 4000 km s$^{-1}$ provides a wall that
may decelerate the faster SN ejecta passing the reverse shock.
Alternatively, some of the extra luminosity of (1--2) $\times10^6$
L$_{\odot}$ needed for the broad absorption features seen on days
120--170 could be supplied, hypothetically, by radioactive $^{56}$Co
decay from $\la$2.5 M$_{\odot}$ of $^{56}$Ni.  Unfortunately,
SN~2006gy was aligned too closely with the Sun after day 240 and we
were no longer able to observe it until the later data reported by
Smith et al.\ (2008b), when it had already faded considerably.  See
Miller et al.\ (2009) for additional discussion of the late-time data.

\begin{figure}
\epsscale{0.99}
\plotone{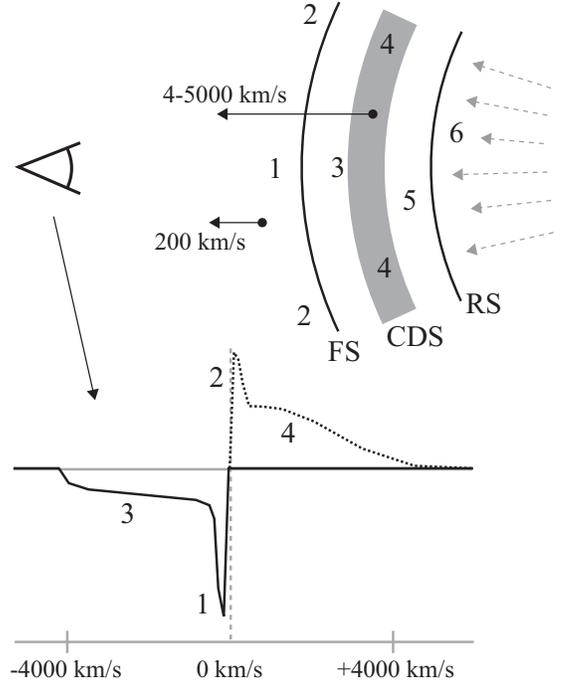}
\caption{Sketch of the emitting geometry in SN~2006gy that leads to
  the formation of its unusual spectral line profiles.  FS, RS, and
  CDS refer to the forward shock, reverse shock, and cold dense shell,
  respectively.  Narrow absorption (1) occurs in the pre-shock gas
  near our line of sight, while narrow emission (2) occurs in other
  parts of the ionized pre-shock gas.  The broad emission component
  (4) is emitted by cooling gas in the CDS, while the broad
  blueshifted absorption (3) occurs somewhere between the FS and the
  continuum photosphere that is within or behind the CDS.  The
  relative strength of components 1, 2, 3, and 4 in a spectral line
  depends on the ionization level of the observed species, its
  relative abundance, and the temperature and density of the gas where
  it is emitted, so not all line profiles look the same.  Many
  overlapping profiles produce the peculiar shape of the spectrum of
  SN~2006gy.  At late times when the CDS begins to fragment and
  becomes optically thin (after day $\sim$115), we may start to see
  broad emission features from swept-up ejecta (5) or the freely
  expanding ejecta (6).}
\label{fig:lineprof}
\end{figure}

\subsection{Origin of the Composite Line Profiles: CSM Interaction
  Seen in Absorption}

Here we take a more detailed and qualitative look at the origin of the
unusual spectrum of SN 2006gy during the second stage described above,
days 90--200 when it showed several overlapping complex, peculiar line
profiles.  While SN~1994W and SN~1995G showed similar spectral lines
at earlier epochs, with narrow emission and absorption, they did not
exhibit such prominent broad absorption features that give SN~2006gy
its distinct spectral appearance.

A cartoon of a hypothetical line profile for a single line is shown in
Figure~\ref{fig:lineprof}.  The narrow and broad absorption
(components 1 and 3, respectively) are common to many lines in the
spectrum.  The narrow and broad emission, components 2 and 4, are seen
in some lines and not in others, depending on the ionization state,
temperature, and density of the gas.  For example, (a) both emission
components are absent in the red Si~{\sc ii} doublet, (b) only the
broad emission is present in Fe~{\sc ii} $\lambda$5018 and Ca~{\sc ii}
$\lambda$8622, and (c) both narrow and broad emission are seen in
H$\alpha$.  One can imagine that a series of these line profiles seen
in superposition, having different strengths of absorption, with or
without either emission component, can lead to the high-resolution
spectrum of SN~2006gy seen in Figure~\ref{fig:zoom}.  Whether
self-consistent radiative transfer calculations can account for the
appearance of the spectrum is a more difficult challenge.

Figure~\ref{fig:lineprof} also indicates the likely places where
components 1, 2, 3, and 4 may be produced in a simplified
CSM-interaction scenario (by ``simplified'', we mean that we do not
show the clumping, instabilities in the CDS, or large-scale asymmetry
that are all possible).  Because of their low velocities, the narrow
components 1 and 2 must arise in the pre-shock CSM, either in
absorption along our line of sight (1) or in emission away from our
direct line of sight (2).  Such narrow features are common in spectra
of SNe~IIn with sufficient resolution.

The broad (actually intermediate-width) emission component (4), which
is a common characteristic of SNe~IIn, is formed in the CDS expanding
at $\sim$4000 km s$^{-1}$ during this time interval (days 90--200).
We showed that the intermediate-width H$\alpha$ emission component can
be approximated as a symmetric composite Gaussian with two components
of roughly equal intensity, one with FWHM = 5200 km s$^{-1}$ and one
with FWHM = 1800 km s$^{-1}$.  While these distinct velocities are
somewhat artificial, this emission component likely arises as a result
of the intrinsic kinematic line profile plus broadening of the line
wings due to electron scattering within the optically thick CDS
itself.  Note that electron scattering within the line-forming region
in the CDS, as suggested by Dessart et al.\ (2009), is different from
the electron scattering that broadens the wings of narrow lines formed
in the CSM that we discussed earlier (\S 3.3.1; see Chugai 2001).  We
suspect that both are at work in SN~2006gy at different times --- one
in the early-time spectra when the photosphere was outside the shock,
and one during the decline from peak luminosity when the photosphere
and H$\alpha$-emitting region were in the CDS.  The idea that electron
scattering plays a role in forming the broad wings of H$\alpha$ is
supported by the fact that they are unchanging with time, and that the
blue H$\alpha$ emission wing extends to faster speeds (5000--6000 km
s$^{-1}$) than the blue edge of the blueshifted H$\alpha$ absorption
at 4000 km s$^{-1}$.
Different lines, like Fe~{\sc ii} and Ca~{\sc ii}, originate from
different optical depths within the CDS and will therefore be modified
to differing degrees by electron scattering.  Whether different
line-profile shapes are consistent with this hypothesis is another
challenge for detailed calculations.  Dessart et al.\ (2009) have
shown, for example, that even different H~{\sc i} lines have different
line wings for this reason.

The broad blueshifted absorption in SN~2006gy (component 3) is unusual
among SNe~IIn because most exhibit symmetric wings of the
intermediate-width H$\alpha$ line, sometimes even showing enhanced
blueshifted emission because the back side of the SN is blocked by the
optically thick CDS.  The dramatic difference in blueshifted H$\alpha$
wings of SN~2006gy and SN~2006tf is shown in Figure~7 of Smith et al.\
(2008a).  While many SNe~IIn do show blueshifted absorption, it is
usually from narrower components formed in the CSM.  The absorption in
component 3 of the SN 2006gy line profiles is far too broad to arise in
the CSM, and must therefore reside between the FS and the photosphere.
The most likely location is in outer layers of the CDS, as long as the
continuum photosphere is in deeper layers.  The source of the
continuum could, in principle, be a combination of background light
from the underlying SN ejecta or continuum generated in deeper layers
of the CDS which absorb X-rays generated in the reverse shock.  This
last scenario requires that the SN ejecta of SN~2006gy can still be an
important source of kinetic energy, even at late times.

\begin{figure}
\epsscale{1.05}
\plotone{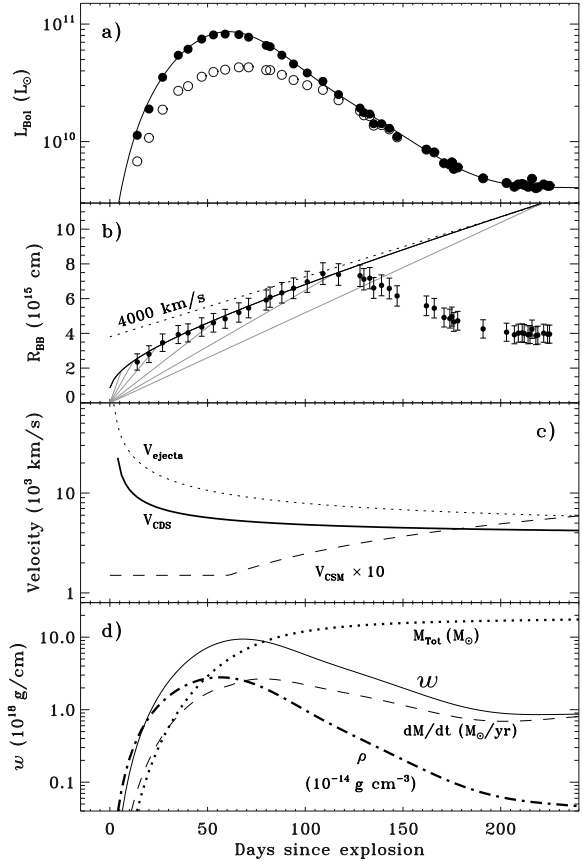}
\caption{A model of the basic parameters of SN 2006gy derived from
  observations in a simplified CSM-interaction formalism.  (a) and (b)
  are the observed luminosity and radius, respectively, from
  Figure~\ref{fig:lc}.  The smoothed curves in both panels are used to
  represent the observed bolometric luminosity and radius of the
  shock, which are needed to derive many other parameters below.  In
  panel (b), the true shock radius continues to increase as $R_{\rm BB}$,
  as described in the text.  The dashed line shows constant-speed
  expansion at 4,000 km s$^{-1}$ for comparison, while the gray lines
  show trajectories for unperturbed SN ejecta in free expansion at 35,
  20, 15, 10, 7.5, and 6 $\times$ 10$^3$ km s$^{-1}$.  (c) The
  corresponding expansion speeds of the cold dense shell, $V_{\rm CDS}$,
  and the SN ejecta that are crossing the reverse shock, $V_{\rm ejecta}$,
  as a function of time.  Values at $t<10$ days are poorly
  constrained.  The dashed line is the assumed speed of the
  progenitor's wind or CSM shell, $V_{\rm CSM}$.  This is taken from
  Figure~\ref{fig:haNarrow}, except that we have set the minimum speed
  at small radii to 150 km s$^{-1}$; this assumption is irrelevant
  except for deriving $\dot{M}$ below.  (If we had allowed $V_{\rm CSM}$
  to continue to decline at small radii, then the derived value of
  $\dot{M}$ before day 70 would be smaller.)  (d) Plotted values are
  the instantaneous values one would derive for the wind density
  parameter, $w$, the progenitor's mass-loss rate, $dM/dt$, the CSM
  density, $\rho$, and the cumulative value of the total CSM mass
  swept up by the shock.  Assumptions for deriving these values are
  explained in the text.}
\label{fig:model}
\end{figure}

\subsection{A Simplified CSM-Interaction Model}

Figure~\ref{fig:model} summarizes several physical aspects of
SN~2006gy interpreted in the format of a standard CSM-interaction
model.  This is not a hydrodynamic model, but rather a set of
parameters derived from observations with a few simplifying
assumptions.  Two key observed values are the SN's bolometric
luminosity and the radius of the CDS which emits the
intermediate-width H$\alpha$ line.  The solid circles in
Figures~\ref{fig:model}a and \ref{fig:model}b are just the bolometric
luminosity and $R_{\rm BB}$ from Figure~\ref{fig:lc}, while the solid
black curves are simple continuous approximations of the data.  We
assume that the true shock radius $R$ continues to increase at roughly
constant speed even though the apparent $R_{\rm BB}$ begins to
decrease around day 115.  The rationale for this is explained in \S
3.2.  Gray lines in Figure~\ref{fig:model}b represent trajectories for
unperturbed SN ejecta.

Figure~\ref{fig:model}c shows three relevant expansion speeds.  At
each time $t$, the speed of the fastest freely expanding SN ejecta
that are reaching the radius of the reverse shock (assumed to be the
same as $R$) is simply $V_{\rm ejecta} = R/t$, shown by the dotted
curve.  The speed of the CDS, $V_{\rm CDS}$, plotted in
Figure~\ref{fig:model} is just the instantaneous value of $\dot{R}$.
For our purposes here, we assume the post-shock cooling zone in this
radiative shock to be very thin, so that $R \approx R_{\rm FS} \approx
R_{\rm RS}$.  The solid line representing $V_{\rm CDS}$ in
Figure~\ref{fig:model}c is consistent with the relatively constant
observed speeds of 4000--5000 km s$^{-1}$ seen in H$\alpha$ from the
time of peak luminosity onward.  Also, these values of $V_{\rm
  ejecta}$ and $V_{\rm CDS}$ are consistent, at least qualitatively,
with expectations of hydrodynamic models of SNe~IIn (e.g., Chugai \&
Danziger 2003).  The dashed line in Figure~\ref{fig:model}c is the
approximate speed of the progenitor's wind or CSM shell, $V_{\rm
  CSM}$, that is being overtaken by the blast wave ($V_{\rm CSM}$ is
multiplied by 10 for display).  This is inferred from observations of
the blue edge of the narrow absorption component of H$\alpha$, shown
in Figure~\ref{fig:haNarrow}.  One exception is that we have set the
minimum speed at small radii to 150 km s$^{-1}$; this assumption is
irrelevant except for the role that $V_{\rm CSM}$ plays in deriving
$\dot{M}$ below.

To calculate the remaining parameters in Figure~\ref{fig:model}d, we
adopt the standard equation for the maximum luminosity that can be
generated by CSM interaction, assuming 100\% efficiency, given by

\begin{equation}
L = \frac{1}{2} w V_{\rm CDS}^3,
\end{equation}

\noindent where $w$ is the wind density parameter $w = 4 \pi R^2
\rho$, or $w = \dot{M} / V_{\rm CSM}$, and $V_{\rm CDS}$ is the value 
for the evolving speed of the CDS derived from $\dot{R}$ above.  The
efficiency may be substantially less than 100\% (a value of 30--50\%
is probably more realistic), which would act to {\it raise} all the
derived values plotted in Figure~\ref{fig:model}d.  The wind density
parameter is plotted here in units of 10$^{18}$ g cm$^{-1}$ (this is
quite a large fiducial value, even for luminous SNe~IIn).  The true
wind density is $\rho = w / 4 \pi R^2$, plotted here in units of
10$^{-14}$ g cm$^{-3}$, and the mass-loss rate of the progenitor star
inferred at a given time is $\dot{M} = w V_{\rm CSM}$.  The value of
M$_{\rm Tot}$ given by the dotted line is the cumulative CSM mass that has
been swept up by the shock and now resides in the CDS.

A key assumption here is that direct radiative cooling from the
post-shock gas supplies the observed instantaneous value of the
luminosity.  We have argued elsewhere that this is problematic on
observational grounds (Paper~I) because of the weakness of H$\alpha$
and X-ray emission compared to the continuum, and have suggested that
some of the peak luminosity may in fact be due to the delayed escape
or diffusion of shock-deposited thermal energy from earlier CSM
interaction.  If true, the net result is that the values of $w$,
$\rho$, and $\dot{M}$ in Figure~\ref{fig:model}d will be 
overestimated around the time of peak luminosity, but severely
underestimated before then.

We see from Figure~\ref{fig:model} that the swept-up mass needed to
provide the luminosity of SN~2006gy in a CSM-interaction scenario is
already 10 M$_{\odot}$ (or perhaps 2--3 times more if the efficiency
in converting kinetic energy to visual light is not 100\%) by the time
of peak luminosity, and almost 20 M$_{\odot}$ by 200 days after
explosion.  This large mass dictates that the dense shell must have
very high optical depths and a long diffusion time.  The $\sim$10
M$_{\odot}$ swept up by the time of peak luminosity matches the value
adopted by Smith \& McCray (2007) in such a model.  It is therefore
likely that the low CSM density before day 50 (i.e.\ at $R \la 4
\times 10^{15}$ cm) plotted in Figure~\ref{fig:model}d is erroneous,
since it stems from the probably false assumption (at early times)
that the observed luminosity directly traces CSM-interaction.  In
reality, high optical depths prevent the shock-deposited thermal
energy from escaping promptly, leading to a value from Eq.\ (1) that
is too low.

The need for higher densities at inner radii (within
4$\times$10$^{15}$ cm) is evident from a separate line of reasoning as
well.  The spectrum at early times (e.g., day 36) arises because the
photosphere is outside the forward shock, since shock luminosity must
diffuse out through the opaque CSM shell.  In the process, multiple
scattering with thermal electrons broadens the narrow H$\alpha$
emission from the photoionized pre-shock gas, causing the symmetric
profiles.  This picture, however, necessitates pre-shock CSM density
on day 36 exceeding that on days 90--100 when the pre-shock CSM is no
longer opaque.  On day 36 we found an optical depth for the pre-shock
CSM of roughly $\tau \approx 15$, whereas on days 90--100 it must have
been $\la$1 because the CDS is seen.  The inferred CSM densities for
days 36 and 90 in Figure~\ref{fig:model}d, however, are comparable,
which cannot be the case.  This discrepancy can be reconciled if we
relax the assumption that the observed luminosity traces the level of
CSM interaction, and is instead due to delayed diffusion at early
times.  Much higher CSM densities, probably above 10$^{-13}$
g~cm$^{-3}$, are required at inner radii of 2$\times$10$^{15}$ cm.
Indeed, CSM densities of 10$^{-13}$ to 10$^{-12}$ g cm$^{-3}$ were
required at these inner radii in models of SN~1994W by Chugai et al.\
(2004), which showed qualitatively similar spectral properties
although at lower luminosity.  SN~2006gy probably requires even higher
densities at these inner radii.

In any case, by late times (200--250 days after explosion,) the total
CSM mass that has been swept up into the CDS by the shock is 18--19
M$_{\odot}$.  This agrees with the value predicted in the models of
Woosley et al.\ (2007) for a pulsational pair-instability ejection.
With a final coasting velocity of $\sim$4000 km s$^{-1}$, the kinetic
energy remaining in the CDS when the strongest CSM-interaction phase
has ended is then at least (2.9--3.0) $\times 10^{51}$ erg (it may be
considerably more if significant mass is added by the SN ejecta
entering the reverse shock).  Combined with the integrated radiated
energy of $E_{rad} =$ (2.3--2.5) $\times 10^{51}$ erg (\S 3.2), the
initial kinetic energy of SN~2006gy was more than $5 \times 10^{51}$
erg and the efficiency in converting energy into visual light was
$<$50\%.  This efficiency, in turn, suggests SN ejecta that were at
least as massive as the swept-up CSM, implicating a very massive
progenitor star.

Qualitatively, the CSM-interaction scenario we suggest for SN~2006gy
is actually quite analogous to scaled-up models proposed for SN~1994W
and SN~1995G by Chugai et al.\ (2004) and Chugai \& Danziger (2003),
respectively.  In both cases, the authors proposed that the observed
luminosity was due to a combination of internal energy leakage from an
extended pre-SN shell, plus subsequent luminosity from CSM
interaction.  This is qualitatively identical to our proposed model
for SN~2006gy, where the prime luminosity source at early times is the
slow diffusion of shock-deposited energy in an unbound opaque CSM
envelope ejected $\sim$8 yr before the SN (see Smith \& McCray 2007),
and data at late times show an increased contribution from the
post-shock shell.  In the case of SN~1994W the pre-SN shell was
ejected 1.5~yr before explosion in a $2 \times 10^{48}$ erg event
(Chugai et al.\ 2004), and for SN~1995G it was ejected 8~yr before in
a $6 \times 10^{48}$ erg event (Chugai \& Danziger 2003).  This is
reminiscent of the SN~2006gy pre-SN outburst, which occurred
$\sim$8~yr before with $\ga 10^{49}$ erg.  Both SN~1994W and SN~2006gy
showed clear evidence that the strength of CSM interaction plummeted
as the blast wave overtook the outer boundary of this CSM shell.

The primary difference between SN~2006gy and the other two events is
that, quantitatively, SN~2006gy is a much more extreme case than
SNe~1994W and 1995G, {\it with 20--50 times times more mass ejected in
  its precursor event}, and a more energetic SN explosion.
Consequently, diffusion through the opaque CSM shell made a more
important contribution to the total luminosity as SN~2006gy remained
fully optically thick (i.e., $R_{\rm BB}$ was equal to the true shock
radius) out to much larger distances.  This is why SN~2006gy had a
smaller H$\alpha$ equivalent width than the other two events (see
Figures~\ref{fig:comp} and \ref{fig:haEW}).  Qualitatively, though,
all three SNe showed similar absorption features in their spectra
(Fig.~\ref{fig:comp}).  As discussed above, some other SNe~IIn like
SN~1998S and SN~2005gl passed through qualitatively similar phases
that, however, lasted only a short time, implying that they are very
slimmed-down versions of similar phenomena.


We have not invoked a high degree of clumping in the CSM of SN~2006gy,
because clumping allows the blast wave to propagate relatively
unimpeded through the lower-density interclump regions (i.e., the
shock remains fast), it facilitates the escape of copious X-rays, and
it allows an observer to see radiation from the underlying SN
photosphere.  These traits are not observed in SN~2006gy.  This is a
key difference between SN~2006gy and models for less luminous (but
X-ray bright) SNe~IIn like SN~1988Z and SN~2005ip, where a clumped CSM
was invoked for these reasons (Chugai \& Danziger 1994; Smith et al.\
2009a).  This is also a difference between our model for SN~2006gy and
that of Agnoletto et al.\ (2009), who assumed strong clumping.

\subsection{Evaluation of the PISN Hypothesis}

In Paper~I we noted that the extraordinary peak luminosity of
SN~2006gy requires exceptional physical conditions unlike those of previous
SNe. If radioactivity is the luminosity source, then an unusually
large mass of $^{56}$Ni near 10 M$_{\odot}$ would be required, which
in turn would necessitate a PISN (e.g., Barkat et al.\ 1967; Rakavy \&
Shaviv 1967; Bond et al.\ 1984; Heger \& Woosley 2002).  If, on the
other hand, CSM interaction were the engine, then a surprisingly large
mass of CSM would be needed, comparable to the 10--20 M$_{\odot}$ of
H-rich material ejected in observed giant LBV eruptions such as the 1843
event of $\eta$ Carinae (Paper~I; Smith et al.\ 2003) or some event
like theoretical pulsational pair-instability eruptions (Woosley et
al.\ 2007).  In Paper~I we were unable to choose between those two
possibilities because both presented challenges, as noted in the
introduction.

After considering the spectral evolution of SN~2006gy in detail, we
find that most of its physical and morphological properties can be
interpreted in the context a two-component CSM-interaction model.  The
apparent contradictions of a simple CSM-interaction model that were
noted in Paper~I can be reconciled with observations by pushing the
physical conditions --- in particular the optical depth and CSM mass
--- to extremes.  High optical depths lead to a situation where the
luminosity of SN~2006gy is powered by a two-component engine:
radiative diffusion from $\sim$10$^{51}$ erg of shock-deposited
thermal energy dominates at early times, while an increasing fraction
at later times is supplied by the normal scenario of direct radiative
post-shock cooling.  This two-component CSM interaction engine seems
to alleviate the need for an internal energy source such as
radioactive decay from 10 M$_{\odot}$ of $^{56}$Ni because it
reconciles the high CSM interaction luminosity with the seemingly
paradoxical weakness of CSM interaction diagnostics like H$\alpha$ and
X-ray emission.

Two final points are worth emphasizing, however.  First, the extremely
dense and massive CSM required for SN~2006gy does seem to be quite
well explained by the predictions of Woosley et al.\ (2007) for a {\it
  pulsational} pair-instability ejection.  This event has the same
underlying instability, but results in a nonterminal precursor
explosion because the energy generation is insufficient to completely
unbind the star.  It is expected to occur for stars with initial
masses in the range 95--135 M$_{\odot}$, just shy of the extremely
high masses required for a true PISN (see Heger et al.\ 2003).
Second, the late-time (day 200--240) luminosity of SN~2006gy is still
a problem, and it remains difficult to decisively rule out some
contribution from radioactive decay.  The late-time luminosity is
sustained even though diagnostics of the optically thin CSM
interaction (H$\alpha$, Fe~{\sc ii}, Ca~{\sc ii}) plummet.  This
luminosity at 200-240 days cannot be mostly due to a light echo (which
{\it does} dominate 1 yr later; Smith et al.\ 2008b; Miller et al.\
2009b) because the optical spectrum is different than at peak, and
because of the strenght of broad absorption features in teh sepctrum
(see \S 3.7).

\subsection{Implications for SN~2005ap and SN~2008es}

The high peak luminosity of SN~2006gy has been exceeded by two recent
SNe, SN~2005ap (Quimby et al.\ 2007) and SN~2008es (Miller et al.\
2009a; Gezari et al.\ 2009).  The interesting twist, though, is that
neither of these had a Type IIn spectrum, unlike all the luminous
runners-up.  It is therefore not obvious that their luminosity is
generated by CSM interaction.

In this paper we have found that SN~2006gy exploded in a dense
two-component CSM, with a massive opaque inner shell within a radius
of (2--4) $\times 10^{15}$ cm, plus a more extended circumstellar
envelope with an outer boundary of roughly 10$^{16}$ cm.  We have
concluded that it was largely the diffusion of radiation from
interaction with the inner opaque shell that powered its rise to peak
luminosity, as proposed by Smith \& McCray (2007), while the continued
CSM interaction with the more extended shell powered some of the
continuum luminosity at late times and determined the H$\alpha$
luminosity, soft X-ray luminosity, and unusual absorption and 
emission-line profiles.  Photoionization of this outer shell produced the
narrow emission-line components in the spectrum.  If this outer shell
had not been present, however, SN~2006gy would not have had a Type IIn
spectrum.

The outer radius of the outer shell of SN~2006gy at $\sim 10^{16}$ cm was
determined primarily by the maximum velocity of the precursor mass
ejection and the fact that it occurred 8~yr before explosion.  If that
ejection episode had less mass and energy, or if it had occurred a
shorter time before the final SN, the resulting CSM envelope may have
been less massive and/or more compact.  By varying the physical
properties of the precursor ejection, we can understand the variety of
observed properties in the most luminous SNe.  A lower CSM mass would
have resulted in faster final expansion speeds due to conservation of
momentum. The lower CSM mass would also cause a faster diffusion time
--- and hence, more rapid rise and decay times --- for interaction
with the opaque shell.  Without the more extended shell, there would
have been no ongoing optically thin CSM interaction to drive the Type
IIn spectrum.  Thus, with a modified set of pre-SN conditions, one can
see how a similar type of CSM interaction could power both SN~2005ap
and SN~2008es.  We therefore consider it unlikely that these events
were true PISNe.  Their properties, however, do seem consistent with
pre-SN shell ejections triggered by a pulsational pair instability, or
perhaps some other as-yet unidentified explosive instability.

\subsection{The Precursor Event: LBV Eruption or the Pulsational Pair 
Instability?}

In Paper~I and in this work, we have noted a number of similarities
between the inferred physical properties of the mass ejection that 
occurred 8~yr before SN 2006gy and the observed properties
of giant LBV eruptions like the 1843 event of $\eta$ Carinae.  The
most important of these are the total ejected mass of 10--20
M$_{\odot}$ (Smith et al.\ 2003), launched at speeds of 100--600 km
s$^{-1}$ with a Hubble-like velocity law indicating a brief event
(Smith 2006a), and the H-rich composition.  We have also noted that
these properties, and the fact that the outburst occurred within a
decade of the SN, seem equally consistent with a precursor event
caused by the pulsational pair instability (Woosley et al.\ 2007).

We wish to emphasize that {\it these two scenarios are not necessarily
  mutually exclusive}.  One is observational (the comparison to
observed LBV eruptions like $\eta$ Car), and the other is a
theoretical idea (the pulsational pair instability).  We cannot rule
out the possibility that they are, in fact, the same phenomenon, since
we still do not know what triggered the 1843 outburst of $\eta$
Car.  After all, most of the observed properties of $\eta$
Car's giant eruption are consistent with expectations of
the pulsational pair instability, except that $\eta$ Car has not
yet met its final demise even though it suffered similar previous
outbursts $\sim$1000 yr ago.  Judging by the additional massive
$\sim$10--20 M$_{\odot}$ dust shell required for the late-time IR light 
echo of SN 2006gy (Smith et al.\ 2008b; Miller et al.\ 2009b), it also
appears to have suffered a previous shell ejection event $\sim$1500~yr
ago.  Heger \& Woosley (2002) have noted that in some cases, the time
between successive pair pulsations can reach as much as $\sim$1000~yr.
Furthermore, recent observations of $\eta$ Car have shown that it even
had an explosive component to the event that created a fast
($\sim$5000 km s$^{-1}$) blast wave out ahead of the most massive
material (Smith 2008).

\subsection{Implications for Massive-Star Evolution: The Role of LBVs?}

Regardless of whether SN 2006gy was a PISN or instead a sequence of pre-SN
shell ejections caused by pair pulsations or some other mechanism, we
argued in Paper~I that the extreme energy and mass budgets make it
difficult to avoid the requirement that SN~2006gy marked the death of
a very massive star.  The work described here reinforces this view:
we find that the amount of CSM required is $\sim$20 M$_{\odot}$, and
that the mass of SN ejecta must be comparable in order to yield the
final efficiency of about 50\% in converting kinetic energy to light.
Furthermore, our previous study of the late-time IR emission (Smith et
al.\ 2008b) suggested a dust shell with a total gas mass of at least
10 M$_{\odot}$ ejected 1000--2000 yr ago.  This mass budget requires
that within $\sim$10$^3$~yr of its final explosion, the progenitor star
still retained no less than 40--50 M$_{\odot}$.

A key point to recall is that SN~2006gy occurred in a galaxy of
roughly solar metallicity, so the progenitor star would have shed
significant mass during its lifetime and its {\it initial} mass was
probably well above 100 M$_{\odot}$.  This is consistent with models
for SN~2006gy as a pulsational pair event (Woosley et al.\ 2007),
which involved a star with an initial mass of 110 M$_{\odot}$.
(Similar requirements appear to be necessary for SN~2006tf, but the
mass budgets are more relaxed for SNe~2005ap and 2008es.)

Therefore, it seems impossible that SN~2006gy was a thermonuclear
(Type Ia) SN that exploded in a dense H-rich CSM to produce a Type~IIn
spectrum, as had been suggested for SNe~2002ic and 2005gj (Hamuy et
al.\ 2003; Chugai \& Yungelson 2004; Aldering et al.\ 2006; Prieto et 
al.\ 2007), and possibly also SNe~1997cy and 1999E (Germany et
al.\ 2000; Turatto et al.\ 2003; Rigon et al.\ 2003).  These had been
the most luminous SNe known until SN~2006gy was discovered.  We have
noted that in several {\it observed} respects, SN~2006gy can be
regarded as a more extreme case of this class of objects.  It
therefore seems possible that not all of them are SNe~Ia in dense
CSM environments either, supporting the view of Benetti et al.\
(2006).  Trundle et al.\ (2008) have also expressed doubt that
SN~2005gj was a Type~Ia event because of similarities between its
pre-shock CSM and the winds of LBV stars.

If SN~2006gy really was the death of a very massive star, it
challenges the standard paradigm that all massive stars at near-solar
metallicity should shed their H envelopes and explode as Wolf-Rayet
(WR) stars (Conti 1976; Chiosi \& Maeder 1986), yielding SNe of 
Type~Ib or Ic and possibly gamma-ray bursts (e.g., Langer et al.\ 1994; 
Maeder \& Meynet 1994; Heger et al.\ 2003; Woosley \& Heger 2006).
Instead, it requires that some extremely massive stars are able to
retain substantial H envelopes up until or immediately before the time
of core collapse.

This brings to mind the recent recognition that steady line-driven
mass-loss rates of O-type stars and WR stars are lower than those
included in most models.  The winds are clumped, which lowers the
mass-loss rate one derives from observations (see Bouret et al.\ 2005;
Fullerton et al.\ 2006).  Lower mass-loss rates for the 3--5 Myr on
the main sequence could substantially increase a star's final mass.
Smith \& Owocki (2006) pointed out that LBV eruptions may help make up
some of the deficit, but if insufficient, the star can explode as an
LBV with part of its H envelope still intact.

We have argued here and elsewhere (Smith et al.\ 2007, 2008a, 2008b;
Smith 2006b) that the extreme progenitor mass-loss rates required to
produce the dense CSM around luminous SNe~IIn may suggest a link to
LBV-like instability in the progenitor stars.  LBVs in eruption are
the {\it only} stars that are observed to have mass-loss rates of the
correct order to explain more luminous SNe~IIn.  Furthermore, there is
empirical evidence that the mass shedding by luminous SN~IIn
progenitors is not due to steady winds, but is instead from episodic
bursts of mass loss like LBV eruptions. Outer shells around at least
some SNe, such as the dusty IR echo shell around SN~2006gy (Smith et
al.\ 2008b), imply that these eruption episodes are recurrent on a
timescale of $\sim$10$^3$ yr, like LBVs.  (We have also noted,
however, that some of the less luminous SNe~IIn, such as SN~2005ip,
can potentially be explained with the most extreme cases of RSG mass
loss, so not {\it all} SNe~IIn necessarily come from LBVs; see Smith
et al.\ 2009a, 2009c.)

The possibility of a LBV/SN~IIn connection, inferred indirectly from
pre-SN CSM properties, has been confirmed by the only case so far
where a SN~IIn progenitor star has been identified directly in
pre-explosion images.  In SN~2005gl it was found that the progenitor
did in fact resemble a very luminous LBV-like star with high
luminosity (Gal-Yam et al.\ 2007, Gal-Yam \& Leonard 2009).  Gal-Yam
\& Leonard found that the progenitor had a temperature consistent with
LBVs (i.e., a blue to yellow supergiant, not a RSG) and a very high
luminosity of $\sim10^6$ L$_{\odot}$.  They also demonstrated that
this star vanished after SN~2005gl faded.  If the progenitor
luminosity is the true luminosity of the quiescent star (and not, for
example, the enhanced luminosity caught during a giant eruption
precursor event), then a very massive progenitor with initial mass
above $\sim$50 M$_{\odot}$ is required. To explain the observed level
of CSM interaction in SN~2005gl, Gal-Yam \& Leonard (2009) find that the
precursor shell ejection event had a mass-loss rate of $\sim$0.03
M$_{\odot}$ yr$^{-1}$, similar to the 1600 AD\ giant eruption of the
famous LBV star P Cygni (Smith \& Hartigan 2006).  The fact that
SN~2005gl itself was not a particularly high-luminosity event compared
to other SNe~IIn underscores the truly astonishing set of conditions
that led to SN~2006gy.  The progenitor of SN~2006gy must have been an
extreme case of an already extraordinary class of stars.

There are several other circumstantial indications that LBVs may be
related to SNe as well, mainly from inferred properties of the H-rich
pre-shock CSM (Salamanca et al.\ 2002; Kotak \& Vink 2006; Smith
2006b, 2007; Smith et al.\ 2007, 2008a, 2008b; Trundle et al.\ 2008).
One argument made by these authors is that the speed and possible
variability of the pre-shock CSM is close to the outflow speeds of a
few hundred km s$^{-1}$ observed in LBVs, while being much faster than
RSG winds and slower than WR winds.  This is true for SN~2005gl as
well (see the Appendix).  Wind speeds alone are not conclusive, since
radiation from the SN itself may accelerate pre-shock CSM to these
speeds.  This could, in principle, cause the two-component absorption
features in SN~2005gj that Trundle et al.\ (2008) showed to be very
similar to H$\alpha$ profiles in the LBV star AG~Carinae.  While not
conclusive on their own, the pre-shock speeds are nevertheless
consistent with the LBV interpretation.  They can be taken together
with the large CSM masses needed to power the luminosity (discussed in
this work) and the similarity between LBV nebulae and the pre-SN
nebula around SN~1987A (Smith 2007) to support the view that in some
cases, LBV episodes may indeed precede SNe.


If other SNe~IIn are also very massive stars that retain their H
envelopes until just before their death, it suggests a picture where
these stars enter a phase of instability shortly before explosion that
triggers the ejection of much or all of their remaining H envelopes.
The underlying SN explosion, which, hypothetically, may otherwise have
been a Type Ia, Ib/c, IIb, II-P, or II-L event, expands into that
recently ejected massive H envelope to produce a SN~IIn. (In this
regard, it may be more appropriate to consider the Type IIn
classification as a phenomenon rather than a class of explosions.)
One might expect a very massive star nearing core collapse to be
unstable if it has retained part of its H envelope, because its outer
H envelope would be dangerously close to, or may substantially exceed,
the classical Eddington limit. The attempt to understand this
phenomenon is likely to benefit from continued study of the physics of
giant LBV eruptions and theoretical investigations of super-Eddington
winds (e.g., Owocki et al.\ 2004; van Marle et al.\ 2009a), even though
it is not yet clear if the two phenomena have the same underlying
trigger.  Such pre-SN instability might be exacerbated by additional
luminosity from final burning stages.  A star could therefore be
subject to periodic and sudden bursts of mass loss as it tries to shed
its massive H envelope.  These clues suggest that the final years of
very massive stars include sporadic events that essentially {\it
  determine} the end fate of the star.  Significant revision is
therefore needed to correctly model and predict the fates of the most
massive stars beyond our speculation noted here, including dynamical
events like LBV eruptions, pulsational pair ejections, or other
instabilities associated with late phases.

\acknowledgments
\scriptsize

N.S.\ acknowledges relevant discussions with S.\ Woosley and A.\
Heger.  We thank the following observers who assisted with runs at
Lick Observatory: M.\ Ganeshalingam, M.\ Moore, D.\ Pooley, S.\
Rodney, T.N.\ Steele, and D.S.\ Wong.  Some of the data presented
herein were obtained at the W. M.\ Keck Observatory, which is operated
as a scientific partnership among the California Institute of
Technology, the University of California, and the National Aeronautics
and Space Administration (NASA); the observatory was made possible by
the generous financial support of the W. M. Keck Foundation.  We wish
to extend special thanks to those of Hawaiian ancestry on whose sacred
mountain we are privileged to be guests.  We are grateful to the
staffs at the Lick and Keck Observatories for their dedicated
services.  KAIT and its ongoing operation were made possible by
donations from Sun Microsystems, Inc., the Hewlett-Packard Company,
AutoScope Corporation, Lick Observatory, the National Science
Foundation (NSF), the University of California, the Sylvia \& Jim
Katzman Foundation, and the TABASGO Foundation. This research was
supported by NSF grant AST--0607485, the TABASGO Foundation, and
NASA/{\it HST} grant GO-10877 from the Space Telescope Science
Institute (STScI), which is operated by AURA, Inc., under NASA
contract NAS 5-26555.  N.S.\ was partially supported by grants GO-10241
and GO-10475 from STScI.

{\it Facilities:} Keck I (LRIS, LRISp), Keck II (DEIMOS), Lick 3~m
(Kast).


\appendix

\begin{figure}
\epsscale{0.6}
\plotone{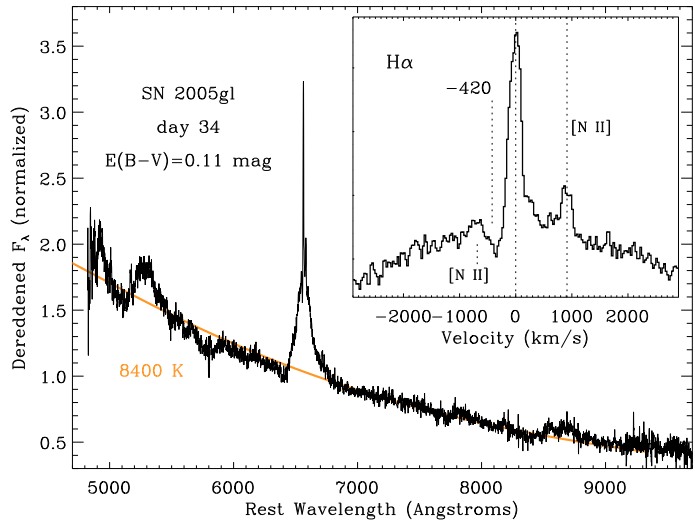}
\caption{The day 34 spectrum of SN~2005gl obtained with DEIMOS, 
  dereddened by $E(B-V) = 0.11$ mag.  An 8500~K blackbody
  is shown for comparison (gray; orange in the online edition).  The
  inset shows the narrow H$\alpha$ profile.  The dashed vertical lines
  indicate the pre-shock speed of 420 km s$^{-1}$ estimated by Gal-Yam \&
  Leonard (2009), as well as the expected location of [N~{\sc ii}]
  lines.}
\label{fig:05gl}
\end{figure}

SN~2005gl is a special case among SNe~IIn because it is the only one,
so far, for which a progenitor star has been identified in
pre-explosion images.  Gal-Yam et al.\ (2007) and Gal-Yam \& Leonard
(2009) have already analyzed the spectral evolution of this SN,
including spectra obtained on days 8, 58, and 87 after discovery.  In
this Appendix, we present an additional unpublished spectrum of
SN~2005gl because our higher-resolution data provide new information
on the progenitor's wind speed, and are therefore pertinent to our
study of SN~2006gy through the LBV/SN~IIn connection.

We observed SN~2005gl using DEIMOS/Keck~II on 2005 November 8, roughly
34 days after discovery.  This is between the first and second epochs
of spectra published by Gal-Yam et al.\ (2007), when the SN
transitioned from one where Type IIn features dominated the spectrum,
to one that looked more like a Type II-P.  Indeed, if we adopt a
redenning correction of $E(B-V) = 0.11$ mag (Gal-Yam et al.\ 2007), we
find a characteristic continuum temperature of $\sim$8500~K
(Figure~\ref{fig:05gl}) that is intermediate between the value of
13,000~K on day 8 and a lower temperature around 6000--6500~K on day
58.  Except for the narrow component, the spectrum is consistent with
those of SNe~II-P at comparable epochs, with a half-width at zero
intensity (HWZI) for the red wing of the broad H$\alpha$ component of
$\sim10^4$ km s$^{-1}$.

Of particular interest is that our spectrum has higher dispersion ($R
= \lambda/\Delta\lambda \approx 3000$) than previous data, resolving
the narrow emission component and revealing the presence of narrow
P~Cygni absorption. Gal-Yam \& Leonard (2009) used the HWZI of the
narrow H$\alpha$ profile, which was marginally unresolved in their
spectra, to estimate a pre-shock outflow speed of $\sim$420 km s$^{-1}$
(shown by the dashed line in Figure~\ref{fig:05gl}).  Our spectrum,
where the line width is fully resolved, yields a slightly faster FWZI
on the red side of roughly 600 km s$^{-1}$, although the line exhibits
some complex structure.  Similarly, our spectrum shows significant
blueshifted P Cygni absorption out to roughly $-$600 km s$^{-1}$ as
well (this is complicated by [N~{\sc ii}] $\lambda$6548 emission, but
absorption is nevertheless present at these speeds).  Although 600 km
s$^{-1}$ is faster than the 420 km s$^{-1}$ estimated by Gal-Yam \&
Leonard (2009), it supports their claim that the speed is similar to
ejecta of LBVs like $\eta$~Car (Smith 2006).  In this paper and Paper
I, we have discussed the same type of evidence for comparable speeds
in the pre-shock CSM of SN~2006gy and made similar connections to
LBV-like stars.  These connections are reinforced by the speeds
observed in the CSM of SN~2005gl.

\end{document}